\documentclass[]{dakota-article}
\pdfoutput=1

\usepackage{amsfonts,amsmath,graphicx,subfigure,amssymb,color}
\usepackage{algorithm, algorithmic}
\usepackage{epsfig} %% for loading postscript figures

\usepackage{tikz}
\usetikzlibrary{positioning}

\newcommand{\prior}[0]{\pi_{{\Lambda}}^{\text{prior}}}

\newcommand{\post}[0]{\pi_{{\Lambda}}^{\text{post}}}

\newcommand{\tildepost}[0]{\tilde\pi_{{\Lambda}}^{\text{post}}}

\newcommand{\obs}[0]{\pi_{\mathbf{\mathcal{D}}}^{\text{obs}}}

\newcommand{\push}[0]{\pi_{\mathbf{\mathcal{D}}}^{Q(\text{prior})}}

\newcommand{\Q}[0]{\mathcal{Q}}
\newcommand{\D}[0]{\mathbf{\mathcal{D}}}
\newcommand{\muL}[0]{\mu_{{\Lambda}}}
\newcommand{\muD}[0]{\mu_{\mathbf{\mathcal{D}}}}

\newcommand{\pspace}{{\Lambda}}
\newcommand{\dspace}{\mathbf{\mathcal{D}}}
\newcommand{\pmeas}{\mu_{\pspace}}
\newcommand{\dmeas}{\mu_{\dspace}}
\newcommand{\pborel}{\mathcal{B}_{\pspace}}
\newcommand{\dborel}{\mathcal{B}_{\dspace}}
\newcommand{\priormeas}{P_{\pspace}^{\text{prior}}}
\newcommand{\postmeas}{P_{\pspace}^{\text{post}}}
\newcommand{\priordens}{\pi_{\pspace}^{\text{prior}}}
\newcommand{\postdens}{\pi_{\pspace}^{\text{post}}}
\newcommand{\pfpriormeas}{P_{\dspace}^{Q(\text{prior})}}

\newcommand{\pfpriordens}{\pi_{\dspace}^{Q(\text{prior})}}

\newcommand{\obsmeas}{P_{\dspace}^{\text{obs}}}
\newcommand{\obsdens}{\pi_{\dspace}^{\text{obs}}}

\newtheorem{remark}{Remark}

\newtheorem{definition}{Definition}
\title{Optimal Experimental Design Using A Consistent Bayesian Approach}

\def\SNL{Optimization and Uncertainty Quantification Department, Sandia National Labs, Albuquerque, NM}
\def\UCD{Department of Mathematical and Statistical Sciences, University of Colorado Denver, Denver, CO 80202}

\def\mycorauthor{J.D. Jakeman}
\corauthor{\mycorauthor}
\coremail{jdjakem@sandia.gov}
\def\mytitle{Optimal Experimental Design Using A Consistent Bayesian Approach}
\title{\mytitle}
\date{\today}
\author{S. Walsh\thanks{\UCD}, T.~Wildey\thanks{\SNL}, J.~Jakeman\samethanks[2] }
\funding{Sandia National
Laboratories is a multimission laboratory managed and operated by
National Technology and Engineering Solutions of Sandia, LLC., a
wholly owned subsidiary of Honeywell International, Inc., for the
U.S. Department of Energy's National Nuclear Security Administration
under contract DE-NA-0003525}
\shortauthor{Walsh, Wildey, Jakeman}
\shorttitle{OED using a consistent Bayesian approach}

\begin{document}
\pagestyle{myheader}
\maketitle

\begin{abstract}
We consider the utilization of a computational model to guide the optimal acquisition of experimental data
to inform the stochastic description of model input parameters.
Our formulation is based on the recently developed {\em consistent} Bayesian approach for solving
stochastic inverse problems which seeks a posterior probability density that is consistent
with the model and the data in the sense that the push-forward of the posterior (through the computational model)
matches the observed density on the observations almost everywhere.
Given a set a potential observations, our optimal experimental design (OED) seeks the observation,
or set of observations, that maximizes the expected information gain from the prior probability density on
the model parameters.
We discuss the characterization of the space of observed densities and a computationally
efficient approach for rescaling observed densities to satisfy the fundamental assumptions
of the consistent Bayesian approach.
Numerical results are presented to compare our approach with existing OED methodologies
using the classical/statistical Bayesian approach and to demonstrate our OED on
a set of representative PDE-based models.
\end{abstract}

%%%%%%%%%%%%%%%%%%%%%%%%%%%%%%%%%%%%%%%%%%%%%%%
%%%%%%%%%%%%%%%%%%%%%%%%%%%%%%%%%%%%%%%%%%%%%%%
%%%%%%%%%%%%%%%%%%%%%%%%%%%%%%%%%%%%%%%%%%%%%%%
%%%%%%%%%%%%%%%%%%%%%%%%%%%%%%%%%%%%%%%%%%%%%%%
\section{Introduction} \label{sec:intro}
Experimental data is often used to infer valuable information about parameters for
models of physical systems.
However, the collection of experimental data can be costly and time consuming.
For example, exploratory drilling can reveal valuable information about subsurface
hydrocarbon reservoirs, but each well can cost upwards of tens of millions of US dollars.
In such situations we can only afford to gather some limited number of
experimental data, however not all experiments provide the same amount
of information about the processes they are helping inform.
Consequently, it is important to design experiments in an optimal way,
i.e., to choose some limited number of experimental data to maximize the
value of each experiment.

The first experimental design methods employed mainly heuristics,
based on concepts such as space-filling and blocking, to select field
experiments~\cite{Cox_R_Book_2000,Fisher_Book_1966,Pazman_book_1986,Pukelsheim_book_1993,Ucinski_book_2005}.
While these methods can perform well in some situations, these methods can be
improved upon by incorporating any knowledge of the underlying physical
processes being inferred or measured.
Using physical models to guide experiment
selection has been shown to drastically improve the cost effectiveness of
experimental designs for a variety of models based on ordinary differential equations \cite{doi:10.1080/10556780410001683078,doi:10.1137/100791063,Bock2013},
partial differential equations \cite{Horesh2010} and differential algebraic equations \cite{Bauer20001}.
When model observables are linear with respect to the model parameters
the alphabetic optimality criteria are often used \cite{Haber2012,doi:10.1137/130933381,0266-5611-24-5-055012}.
For example
$A$-optimality to minimize the average variance of parameter
estimates, $D$-optimality to maximize the differential Shannon
entropy, or $G$-optimality to minimize the maximum variance of model
predictions. These criteria have been developed in both Bayesian and
non-Bayesian settings~\cite{doi:10.1137/140992564,doi:10.1137/130933381,Haber2012,Atkinson_D_book_1992,Chaloner_V_IMS_1995,Long_laplace}.

In this paper we focus attention on Bayesian methods for OED that can be applied to both
linear and nonlinear models~\cite{Loredo_C_SCA_2003, Muller_SD_JASA_2004,
Solonen_HL_JCGS_2012}.
Specifically we pursue OEDs which are optimal for inferring model
parameters on finite-dimensional spaces
from experimental data observed at a set of sensor locations.
%The form of an optimal experimental design is
%dependent on the exact criteria and objective functions used to measure the
%benefit of an experiment.
In the context of OED for inference,
analogues of the alphabetic criterion, for linear models have also been applied to nonlinear
models~\cite{doi:10.1137/140992564, Long_STW_CMAME_2013,0266-5611-26-2-025002,Horesh2010}.
In certain situations,
for example infinite-dimensional problems (random variables are
random fields) or problems with computational expensive
models, OED based upon linearizations of the model response and
Laplace (Gaussian) approximations of the
posterior distribution have been
necessary~\cite{doi:10.1137/140992564, Long_STW_CMAME_2013}.
In other settings non-Gaussian approximations of the posterior have also been
 pursued~\cite{Loredo_C_SCA_2003, Muller_SD_JASA_2004,Solonen_HL_JCGS_2012, Long_laplace}.

%Therefore, we can only seeks to maximize the expected information gain, i.e.,
%the average information gain over all possible realizations of a given set of experimental data.
%The determination of an OED is based largely on the computational model of the physical system.
%With no experimental data available, the physics and sensitivities of the computational model
%greatly impact our definition of the OED.

This manuscript presents a new approach for OED based upon {\em consistent} Bayesian inference, introduced
in \cite{cbayes}. We adopt an approach for OED similar to the
approach in~\cite{huan} and seek an OED that
maximizes the expected information gain from the prior to the
posterior over the set of possible observational
densities. Although our OED framework is Bayesian in nature, this approach is
fundamentally different from the statistical Bayesian methods
mentioned above. The aforementioned Bayesian OED methods use what we
will refer to as the classical/statistical Bayesian approach for
stochastic inference (see e.g., \cite{Stuart_IP_2010}) to
characterize posterior densities that reflects an assumed error model.
In contrast, consistent Bayesian inference assumes a probability density on the observations
is given and produces a posterior density that is
consistent with the model and the data in the sense that the push-forward
of the posterior (through the computational model) matches the
observed density almost everywhere. We direct the interested reader to
\cite{cbayes} for a discussion on the differences between the
consistent and statistical Bayesian approaches.
Consistent Bayesian inference has some connections with measure-theoretic
inference~\cite{BE1}, which was used for OED in~\cite{measure_oed},
but the two approaches make different assumptions and therefore
typically give different solutions to the stochastic inverse problem.

The consistent Bayesian approach is appealing for OED since it can be
used in an offline-online mode.
Consistent Bayesian inference requires
an estimate of the push-forward of the prior, which although expensive
can be computed offline or obtained from archival simulation data.
Once the push-forward of the prior is constructed, the posterior density
can be approximated cheaply.
Moreover, this push-forward of the prior does not depend on the density on the observations
which enables a computationally efficient approach for solving multiple stochastic inverse
problems for different densities on the observations.
This can significantly reduce the cost of computing the expected information gain
if the set of candidate observation is known {\em a priori}.

The main objectives in this paper are to derive an OED formulation using the consistent
Bayesian framework and to present a computational strategy to estimate the
expected information gained for an experimental design.
The pursuit of a computationally efficient approach for coupling our OED method with continuous optimization techniques is an intriguing topic that we leave for future work.
%For the sake of brevity, we do not discuss or pursue efficient optimization strategies to find the OED.
Here, we consider batch design over a
discrete set of possible experiments. Batch design, also known as open-loop
design, involves selecting a set of experiments concurrently such
that the outcome of any experiment does not effect the selection of
the other experiments. Such an approach is often necessary when one
cannot wait for the results of one experiment before starting
another, but is limited in terms of the number of observations
we can consider.

%Finally we remark that in this paper we consider batch design over a
%discrete set of possible experiments. Batch design, also known as open-loop
%design, involves selecting a set of experiments concurrently such
%that the outcome of any experiment does not effect the selection of
%the other experiments. Such an approach is often necessary when one
%cannot wait for the results of one experiment before starting
%another.

%This approach only uses local linear approximations of quantity of interest (QoI) maps and their singular values to approximate the OED.

%{\color{red} TMW: What are the main contributions of this paper?
%I think it is important to show that we can use the consistent Bayes method to find an OED.
%It is also important to highlight the computational advantages in our approach since
%we only need the push-forward of the prior to do all of the inversions.
%Also, I think the discussion on the proper characterization of the infeasible data is important.
%I don't think a discussion on this will be found in the statistical Bayesian literature and we should
%mention why.}

The remainder of this paper is outlined as follows.
In Section~\ref{sec:consitent} we summarize the consistent Bayesian method for solving stochastic inverse problems.
%This method provides a posterior density that is consistent with the model and observed data.
In Section~\ref{sec:the_information} we discuss the information content of an experiment, and present our OED formulation based upon expected information gain.  During the process of defining the expected information gain of a given experimental design,
care must be taken to ensure that the model can predict all of our potential observed data. In Section~\ref{sec:infeasible} we discuss situations for which this assumption is violated and means for avoiding these situations.
Numerical examples are presented in Section~\ref{sec:numerical} and concluding remarks are provided in Section~\ref{sec:conclusion}.

%%% Local Variables:
%%% mode: latex
%%% TeX-master: "../oed_cbayes"
%%% End:

%%%%%%%%%%%%%%%%%%%%%%%%%%%%%%%%%%%%%%%%%%%%%%%
%%%%%%%%%%%%%%%%%%%%%%%%%%%%%%%%%%%%%%%%%%%%%%%
% input the consistent Bayes background
\section{A Consistent Bayes formulation for stochastic inverse
  problems} \label{sec:consitent}
We are interested in experimental designs which are optimal for inferring model
parameters from experimental data. Inferring model parameters for a
single design and realization of experimental data is a fundamental
component of producing such optimal designs.
In this section we summarize the consistent Bayes method for
parametric inference, originally presented in \cite{cbayes}.
Although Bayesian in nature, the consistent Bayesian approach differs
significantly from its classical Bayesian counterpart~\cite{Stuart_IP_2010,Sargsyan_NG_IJCK_2015,Kennedy_O_JRSSSB_2001} which was used for OED in
\cite{huan,Huan2013288,bayesian_review,doi:10.1137/130933381,doi:10.1137/140992564,Long_STW_CMAME_2013,Long_laplace,Huan_2014}.
We refer the interested reader to \cite{cbayes} for a full discussion of these differences.

%%%%%%%%%%%%%%%%%%%%%%%%%%%%%%%%%%%%%%%%%%%%%%%%%%%%%%%%%%%%%%%%%%%%
%%%%%%%%%%%%%%%%%%%%%%%%%%%%%%%%%%%%%%%%%%%%%%%%%%%%%%%%%%%%%%%%%%%%
\subsection{Notation, Assumptions, and a Stochastic Inverse Problem}\label{subsec:problems}
%%%%%%%%%%%%%%%%%%%%%%%%%%%%%%%%%%%%%%%%%%%%%%%%%%%%%%%%%%%%%%%%%%%%
%%%%%%%%%%%%%%%%%%%%%%%%%%%%%%%%%%%%%%%%%%%%%%%%%%%%%%%%%%%%%%%%%%%%

Let $M(Y,\lambda)$ denote a deterministic model with solution $Y(\lambda)$ that is an implicit function of model parameters $\lambda\in\pspace \subset \mathbb{R}^n$.
The set $\pspace$ represents the largest physically meaningful domain of parameter values, and, for simplicity, we assume that $\pspace$ is compact.
In practice, modelers are often only concerned with computing  a relatively small set of quantities of interest (QoI), $\{Q_i(Y)\}_{i=1}^m$, where each $Q_i$ is a real-valued functional dependent on the model solution $Y$.
Since $Y$ is a function of parameters $\lambda$, so are the QoI and we write $Q_i(\lambda)$ to make this dependence explicit.
Given a set of QoI, we define the QoI map $Q(\lambda) := (Q_1(\lambda), \cdots, Q_m(\lambda))^\top:\pspace\to\dspace\subset\mathbb{R}^m$ where $\dspace  := Q(\pspace)$ denotes the range of the QoI map.
% The QoI can often be categorized into two types: observable QoI and prediction QoI.
% An observable QoI may correspond to specific computable quantities that are either currently available or attainable/verifiable from experimental/field data.
% A prediction QoI may correspond to either current values of state variables $Y(\lambda)$ that are unobservable or a particular quantity that occurs either at a future time or with respect to a different model setup with inputs from $\pspace$.
%In the following we will define an inverse problem which uses observed data which can be compared to the simulated QoI in order to infer information on $\pspace$.
%We refer to this as the inverse for inference process to emphasize the impact that formulating and solving a specific type of inverse problem has on the ability to draw inferences.
%Inferences drawn on $\pspace$ may subsequently be used to study the prediction QoI.
%In this paper, we focus on the inverse for inference process and leave the issues associated with the study of prediction QoI to future work.

%\subsection{Defining the stochastic inverse problem}
Assume $(\pspace, \pborel, \pmeas)$ and $(\dspace, \dborel, \dmeas)$ are measure spaces.
We assume $\pborel$ and $\dborel$ are the Borel $\sigma$-algebras inherited from the metric topologies on $\mathbb{R}^n$ and $\mathbb{R}^m$, respectively.
The measures $\pmeas$ and $\dmeas$ are volume measures.
% Similarly let $(\dspace, \dborel, \dmeas)$ using the Borel $\sigma$-algebra $\dborel$ and measure $\dmeas$. We refer to $\pmeas$ and $\dmeas$ as volume measures.
%The role of the volume measures is quite different than the role of a probability measure, but nonetheless proves vital in a variety of stochastic and non-stochastic analyses \cite{:BE1, BE3, Billingsly, Ferenczi1997}.
%In fact, when a probability measure is described in terms of a probability density function, a volume measure is at least implicitly defined, and is often assumed to be the Lebesgue measure, e.g., see the Radon-Nikodym theorem in \cite{Billingsly}.

%We direct the interested reader to \cite{??} for some background information on this important topic.

%We also define a measure space $(\dspace, \dborel, \dmeas)$ using the Borel $\sigma$-algebra $\dborel$ and measure $\dmeas$, which we also refer to as a volume measure.
%There are possibly several choices for $\dmeas$ in terms of either induced, push-forward, volume measures using the map $Q$ or in terms of more ``natural'' choices based on pointwise descriptions of $\dspace$.
%We return to this topic below when discussing densities.

We assume that the QoI map $Q$ is at least piecewise smooth implying that $Q$ is a measurable map between the measurable spaces $(\pspace, \pborel)$ and $(\dspace, \dborel)$.
For any $A\in\dborel$, we then have
\[Q^{-1}(A) = \left\{ \lambda \in \pspace \ | \ Q(\lambda) \in A \right\}\in\pborel, \quad \text{and} \quad Q(Q^{-1}(A))=A.\]
Furthermore, $B \subseteq Q^{-1}(Q(B))$ for any $B\in\pborel$, although in most cases $B\neq Q^{-1}(Q(B))$ even when $n=m$.% (see Figure~\ref{fig:maps} where $n=2$ and $m=1$ for an illustration).
%\begin{figure}[ht]
% \begin{center}
%   %\includegraphics[trim=0cm 4cm 0cm 4cm, clip=true, scale=0.5]{figures/uqmaps.jpg}
%   \includegraphics[width=0.7\columnwidth]{plots/measure-theoretic-set-example.pdf}
% \end{center}
% \caption{Illustration of the mapping $Q:\pspace \to \dspace$.  Note that $Q$ maps $B$ to a particular region in $\dspace$ ($Q(B)$), and while the inverse image of this set, given by $Q^{-1}(Q(B))$, contains $B$, other points in $\pspace$ may also map to $Q(B)$.  (Figure adopted from \cite{cbayes}).}
% \label{fig:maps}
% \end{figure}

Finally, we assume that an observed probability measure, $\obsmeas$, is given on $(\dspace,\dborel)$ and is absolutely continuous with respect to $\dmeas$, which implies it can be described in terms of an observed probability density, $\obsdens$.
The stochastic inverse problem is then defined as determining a probability measure, $P_\pspace$, described as a probability density, $\pi_\pspace$,
such that, the push-forward measure agrees with $\obsmeas$.
We use $P^{Q(P_\pspace)}_\dspace$ to denote the push-forward of $P_\pspace$ through $Q(\lambda)$, i.e.,
\begin{equation*}
P^{Q(P_\pspace)}_\dspace(A) = P_\pspace(Q^{-1}(A)).
\end{equation*}
for all $A\in \dborel$.
Using this notation, a solution to the stochastic inverse problem is defined formally as follows:
\begin{definition}[Consistency]\label{def:inverse-problem}
Given a probability measure $\obsmeas$ on $(\dspace, \dborel)$ that is absolutely continuous with respect $\dmeas$ and admits a density $\obsdens$,
the stochastic inverse problem seeks a probability measure $P_\pspace$ on $(\pspace, \pborel)$ that is absolutely continuous with respect to $\pmeas$ and admits a probability density
$\pi_\pspace$,
such that the subsequent push-forward measure induced by the map, $Q(\lambda)$, satisfies
\begin{equation}\label{eq:invdefn}
P_\pspace(Q^{-1}(A)) = P^{Q(P_\pspace)}_\dspace(A) = \obsmeas(A),
\end{equation}
for any $A\in \dborel$.
We refer to any probability measure $P_\pspace$ that satisfies \eqref{eq:invdefn} as a {\bf consistent} solution to the stochastic inverse problem.
\end{definition}

%We refer to any probability measure that satisfies \eqref{eq:invdefn} as a {\em consistent} solution
%since it is consistent with both the model and the data.
%In Section \ref{sec:solvability}, we provide the necessary and sufficient conditions under which a solution of
%the stochastic inverse problem exists.
Clearly, a consistent solution may not be unique, i.e., there
may be multiple probability measures that are consistent in the sense of Definition~\ref{def:inverse-problem}.
This is analogous to a deterministic inverse problem where multiple sets of parameters may produce the
observed data.
A unique solution may be obtained by imposing additional constraints or structure on the stochastic inverse problem.
In this paper, such structure is obtained by incorporating prior information to construct a unique
Bayesian solution to the stochastic inverse problem.

\subsection{A Bayesian solution to the stochastic inverse problem}\label{subsec:bayesbrief}
Following the Bayesian philosophy \cite{Tarantola}, we introduce a {\em prior} probability measure $\priormeas$ on $(\pspace, \pborel)$ that is absolutely continuous with respect to $\pmeas$ and admits a probability density $\priordens$.
The prior probability measure encapsulates the existing knowledge about the uncertain parameters.
%The topic of choosing an appropriate prior probability measure is important, and while we describe the basic assumptions required of the prior for our purposes below, a full discussion of this topic is beyond the scope of this work.

Assuming that $Q$ is at least measurable, then the prior probability measure on $\pspace$, $\priormeas$, and the map, $Q$, induce a push-forward measure $\pfpriormeas$ on $\dspace$, which is defined for all $A\in \dborel$,
%\begin{equation}\label{eq:pfprior}
%\pfpriormeas(A) = \int_A \pfpriordens d\dmeas = \int_{Q^{-1}(A)} \pfpriordens d\pmeas = \priormeas(Q^{-1}(A)).
%\end{equation}
\begin{equation}\label{eq:pfprior}
\pfpriormeas(A) = \priormeas(Q^{-1}(A)).
\end{equation}

We utilize the following expression for the posterior,
\begin{equation}\label{eq:setbayes}
\postmeas(B) := \begin{cases}
	\priormeas(B) \frac{\obsmeas(Q(B))}{\pfpriormeas(Q(B))}, & \text{if } \priormeas(B)>0, \\
	0, & \text{otherwise},
	\end{cases}
\end{equation}
which we describe in terms of a probability density given by
\begin{equation}\label{eq:postpdf}
 \postdens(\lambda) = \priordens(\lambda)\frac{\obsdens(Q(\lambda))}{\pfpriordens(Q(\lambda))}, \quad \lambda \in \pspace.
\end{equation}
We note that if $\pfpriordens = \obsdens$, i.e., if the prior solves the stochastic inverse problem,
then the posterior density will be equal to the prior density.
%In Section~\ref{sec:Bayes}, we derive \eqref{eq:setbayes} using Bayesian concepts for measureable sets and
%prove that it solves the stochastic inverse problem \eqref{eq:invdefn} and can be
%described as the probability density given in \eqref{eq:postpdf}.

It was recently shown in \cite{cbayes} that the posterior given by
\eqref{eq:setbayes} defines a consistent probability measure using a {\em contour} $\sigma$-algebra.
When interpreted as a particular iterated integral of \eqref{eq:postpdf}, the posterior defines a probability measure on $(\pspace, \pborel)$ in the sense of Definition~\ref{def:inverse-problem},
i.e., the push-forward of the posterior matches the observed probability density.
Approximating the posterior density using the consistent Bayesian approach only requires an approximation
of the push-forward of the prior probability on the model parameters, which is fundamentally a forward
propagation of uncertainty.
While numerous approaches have been developed in recent years
to improve the efficiency and accuracy of the forward propagation of uncertainty using computational models,
in this paper we only consider the most basic of methods, namely Monte Carlo sampling,
to sample from the prior.
We evaluate the computational model for each of the samples from the prior and use
a standard kernel density estimator~\cite{wand1994multivariate} to approximate the push-forward of the prior.

Given the approximation of the push-forward of the prior, we can evaluate the posterior at any point
$\lambda \in \pspace$ if we compute $Q(\lambda)$.
This provides several possibilities for interogating the posterior.
In Section~\ref{sec:simple}, we compute $Q(\lambda)$ on a uniform grid of points to visualize the posterior  after we compute the push-forward of the prior.
This does require additional model evaluations, but visualizing the posterior is rarely required and only useful for illustrative purposes in 1 or 2 dimensions.
More often, we are interested in obtaining samples from the posterior.
This is also demonstrated in Section~\ref{sec:simple} where the samples from the prior
are either accepted or rejected using a standard rejection sampling procedure.
For a given $\lambda$, we compute the ratio $\postdens(\lambda)/(M\priordens(\lambda))$, where $M$ is an estimate of the maximum of the ratio over $\pspace$, and compare this value with a sample, $\eta$, drawn from a uniform distribution on (0,1).  If the ratio is larger than $\eta$, then we accept the sample.
We apply the accept-reject algorithm to the samples from the prior and therefore the samples from the posterior are a subset of the samples used to compute the push-forward of the prior.
Since we have already computed $Q(\lambda)$ for each of these samples, the computational cost to select a subset of the samples for the posterior is minimal.
%On the other hand, this may not generate many samples from the posterior if the prior and posterior are significantly different.
%We are currently developing better sampling strategies to generate samples from the consistent Bayesian posterior.
However, in the context of OED we are primarily interested in computing the information gained from the prior to the posterior which only involves integrating with respect to the prior (see Section~\ref{sec:information}) and does not require additional model evaluations or rejection sampling.

In practice, we prefer to use data that is sensitive to the parameters since otherwise it is difficult to infer useful information about the uncertain parameters.  Specifically, if $m\leq n$ and the Jacobian of $Q$ is defined a.e. in $\pspace$ and is full rank a.e., then the push-forward volume measure $\muD$ is absolutely continuous with respect to the Lebesgue measure \cite{cbayes}.

For the rest of this work we maintain the following assumptions needed to produce a unique consistent solution to the stochastic inverse problem:

{\bf (A1)} We have a mathematical model and a description of our prior knowledge about the model input parameters,

{\bf (A2)} The data exhibits sensitivity to the parameters a.e. in $\pspace$, hence, we use the Lebesgue measure $\mu$ as the volume measure on the data space,

{\bf (A3)} The observed density is absolutely continuous with respect to the push-forward of the prior.

The assumption concerning the absolute continuity of the observed density with respect to the prior is essential to define a solution to the stochastic inverse problem \cite{cbayes}.
While this assumption may appear rather abstract, it simply assures that the prior and the model can predict, with non-zero probability, any event that we have observed.
Since the observed density and the model are assumed to be fixed, this is only an assumption on the prior.

In the remainder of this work, we focus on quantifying the value of these posterior densities.
We use the Kullback-Leibler divergence \cite{kl,renyikl}, to measure the information gained
about the parameters from the prior to the posterior.
We compute the expected information gain of a given set of QoI (a given experimental design),
and then determine the OED to deploy in the field.

%%% Local Variables:
%%% mode: latex
%%% TeX-master: "../oed_cbayes"
%%% End:

%%%%%%%%%%%%%%%%%%%%%%%%%%%%%%%%%%%%%%%%%%%%%%%
%%%%%%%%%%%%%%%%%%%%%%%%%%%%%%%%%%%%%%%%%%%%%%%
%%%%%%%%%%%%%%%%%%%%%%%%%%%%%%%%%%%%%%%%%%%%%%%
%%%%%%%%%%%%%%%%%%%%%%%%%%%%%%%%%%%%%%%%%%%%%%%
\section{The information content of an
  experiment} \label{sec:the_information}
We are interested in finding the OED for
inferring model input parameters. Conceptually, a design is informative
if the posterior distribution of the model parameters is significantly
different from the prior. To quantify the {\it information gain} of a
design we use the Kullback-Leibler (KL) divergence \cite{renyikl}
as a measure of the difference between a prior and posterior distribution.
While the KL divergence is by no means the only way to compare two
probability densities, it does provide a reasonable measure of the
information gained in the sense of Shannon information
\cite{Cover} and is commonly used in Bayesian OED
\cite{Huan2013288}. In this section we discuss how to compute the KL
divergence and define our OED formulation based upon expected information gain over a specific space of possible observed densities.

%%%%%%%%%%%%%%%%%%%%%%%%%%%%%%%%%%%%%%%%%%%%%%%
\subsection{Information gain: Kullback-Leibler divergence} \label{sec:information}
Suppose we are given a description of the uncertainty on the observed
data in terms of a probability density $\obs$.
This produces a unique solution to the stochastic inverse problem
$\postmeas$ that is absolutely continuous
with respect to the Lebesgue measure $\muL$~\cite{cbayes} and admits a probability
density, $\postdens$.
The KL divergence of the posterior from  the prior (information gain), denoted $I_Q$, is given by
\begin{equation}\label{eq:kl}
  I_Q(\prior:\post) := \int_{\pspace}\post\log\Bigg({\frac{\post}{\prior}}\Bigg) d\muL.
\end{equation}
%where we choose the volume measure, $\muL$, to be a scaled Lebesgue measure such that, for the compact $\pspace$, $\muL(\pspace)=1$.  This choice of $\muL$ impacts the scaling of $\prior$ and $\post$ which in turn produces an information gain that is independent of the measure of the parameter space, allowing us to compare $I_Q$ values obtained using different models.
Note that because $\prior$ is fixed, $I_Q$ is simply a function of the posterior
\begin{equation}\label{eq:kl_post}
  I_Q(\prior:\post) = I_Q(\post),
\end{equation}
and from Eq.~\eqref{eq:postpdf} the posterior is a function of the observed density.  Therefore, we write $I_Q$ as a function of the observed density,
\begin{equation}\label{eq:kl_obs}
  I_Q(\post) = I_Q(\obs).
\end{equation}
The observation that $I_Q$ is a function of only $\obs$ allows us to define the expected information gain in Section~\ref{sec:expected} based on a specific space of observed densities.

Given a high dimensional parameter space, it may be computationally infeasible to accurately approximate the integral in Eq.~\eqref{eq:kl}.
For example, a multi-variate normal density with unit variance in 100-dimensions has a maximum value of
$(1/\sqrt{2\pi})^{100} \approx 1\times 10^{-40}$.
However, we may write this integral in terms of densities on the data space evaluated at $Q(\lambda)$ as follows
\begin{eqnarray}
	I_Q(\post) &=& \int_{\pspace}\post(\lambda)\log\Bigg({\frac{\post(\lambda)}{\prior(\lambda)}}\Bigg) d\muL \nonumber\\
	&=& \int_{\pspace}\prior(\lambda)\frac{\obs(Q(\lambda))}{\push(Q(\lambda))} \log\Bigg({\frac{\obs(Q(\lambda))}{\push(Q(\lambda))}}\Bigg) d\muL \nonumber\\
	&=& \int_{\pspace}\frac{\obs(Q(\lambda))}{\push(Q(\lambda))} \log\Bigg({\frac{\obs(Q(\lambda))}{\push(Q(\lambda))}}\Bigg) d\priormeas, \label{eq:kl_dataspace}
\end{eqnarray}
where the second equality comes from a simple substitution using Eq.~\ref{eq:postpdf}.
%and the third equality follows from the Radon-Nikodym theorem \ref{Bogachev}.
Given a set of samples from the prior, we only need to compute the push-forward of the prior in the data space to approximate $I_Q$. 
This observation provides an efficient method for approximating $I_Q$ given a high dimensional parameter space and a low dimensional data space. 
In fact, we found it convenient to use \eqref{eq:kl_dataspace} whenever the prior is not uniform.
In the consistent Bayesian formulation, we evaluate the model at the samples generated from the prior to estimate the push-forward of the prior.
It is a computational advantage to also use these samples to integrate with respect to the prior rather than integrating
with respect to the volume measure which would require additional model evaluations. 
%See Section~\ref{??} for an application in which we utilize Eq.~\ref{eq:kl_dataspace}.

%%%%%%%%%%%%%%%%%%%%%%%%%%%%%%%%%%%%%%%%%%%%%%%
\subsection{A motivating nonlinear system} \label{sec:simple}
\begin{figure}[htbp]
  \begin{center}
    \scalebox{0.3}{\includegraphics{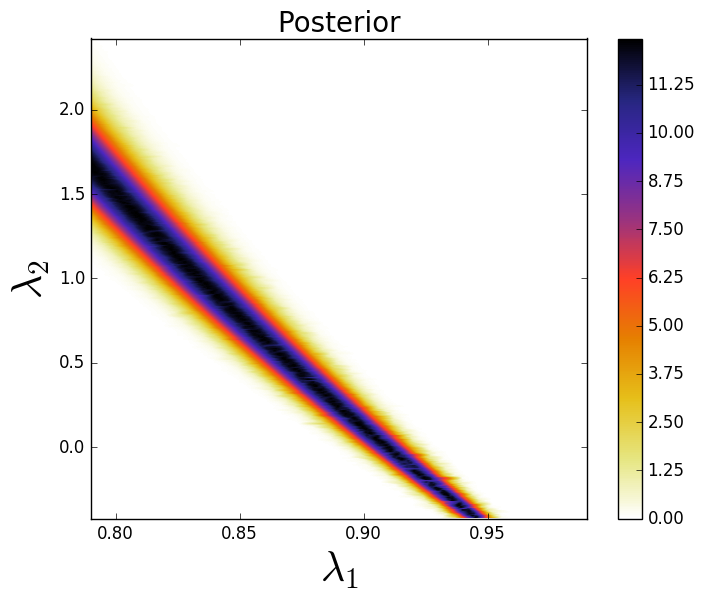}}
    \scalebox{0.3}{\includegraphics{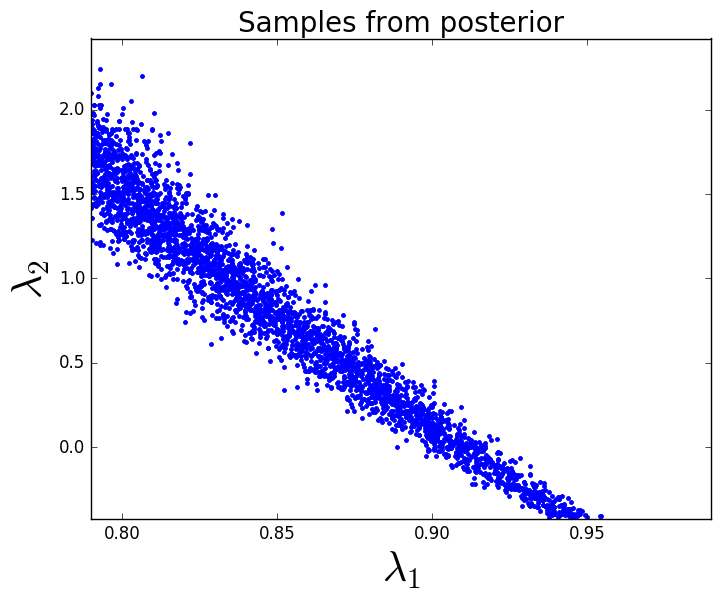}}
    \scalebox{0.3}{\includegraphics{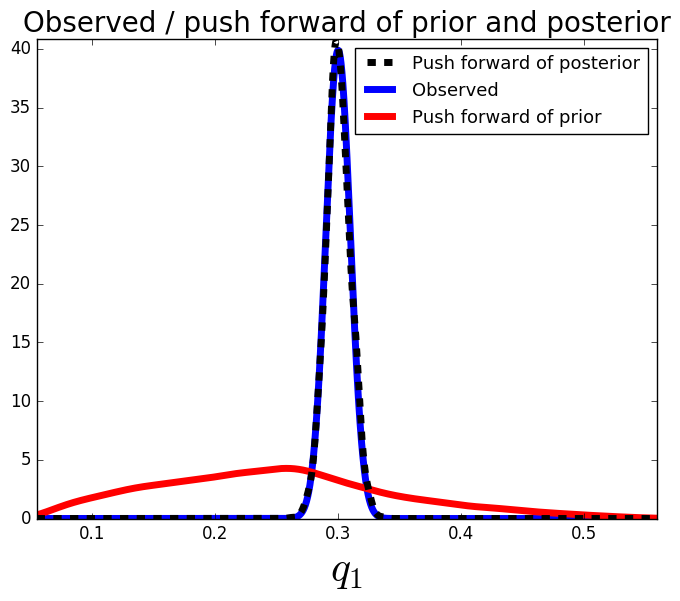}}
    \caption{Approximation of the posterior density obtained using the data $Q_1$ (left) which gives $I_{Q_1}(\obs)\approx 2.015$, a set of samples from the posterior (middle), and a comparison of the observed density on $Q_1$ with the push-forward densities of the prior and the posterior (right).}\label{fig:1stqoi}
  \end{center}
\end{figure}

Consider the following 2-component nonlinear system of equations with two parameters introduced in \cite{BE1}:
\begin{eqnarray*}
  \lambda_1x_1^2 +x_2^2 = 1\\
  x_1^2 - \lambda_2x_2^2 = 1
\end{eqnarray*}
The first QoI is the second component, i.e., $Q_1(\lambda)=x_2(\lambda)$.  The parameter ranges are given by $\lambda_1\in[0.79, 0.99]$ and $\lambda_2\in[1-4.5\sqrt{0.1}, 1+4.5\sqrt{0.1}]$ which are chosen as in \cite{BE1} to induce an interesting variation in the QoI.  We assume the observed density on $Q_1$ is a truncated normal distribution with mean 0.3 and standard deviation of 0.01, see Figure~\ref{fig:1stqoi} (right).

%An analytical solution for the posterior is not available for the consistent Bayes problem, so
We generate 40,000 samples from the uniform prior and use a kernel density estimator (KDE) to construct an approximation to the resulting push-forward density, see Figure~\ref{fig:1stqoi} (right).  Then we use Eq.~\eqref{eq:postpdf} to construct an approximation to the posterior density using the same 40,000 samples, see Figure~\ref{fig:1stqoi} (left), and a simple accept/reject algorithm to generate a set of samples from the posterior, see Figure~\ref{fig:1stqoi} (middle).  We propagate this set of samples from the posterior through the model and approximate the resulting push-forward of the posterior density using a KDE.  In Figure~\ref{fig:1stqoi} (right) we see the push-forward of the posterior agrees quite well with the observed density.
% Notice the properties of this posterior.  Its support lies in a relatively small region of the parameter space, however, it extends over large regions of each of $\lambda_1$ and $\lambda_2$.  The information gain from this posterior is $I_{Q_1}(\obs) = 2.037$.   Next, we consider a different QoI to use in the inverse problem, and compare the properties of its posterior to those we just observed.
Notice the support of the posterior lies in a relatively small region of the parameter space.
%, however, it extends over large regions of each of $\lambda_1$ and $\lambda_2$.
The information gain from this posterior is $I_{Q_1}(\obs)\approx 2.015$.

\begin{figure}[htbp]
  \begin{center}
    \scalebox{0.3}{\includegraphics{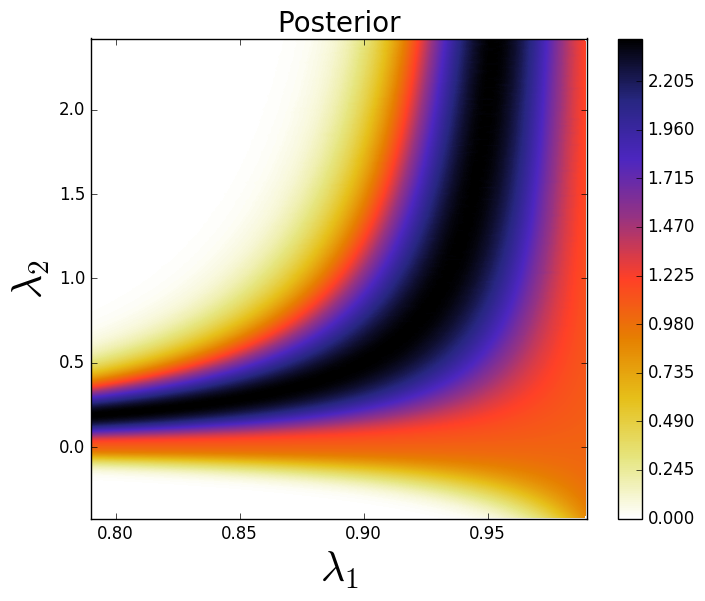}}
    \scalebox{0.3}{\includegraphics{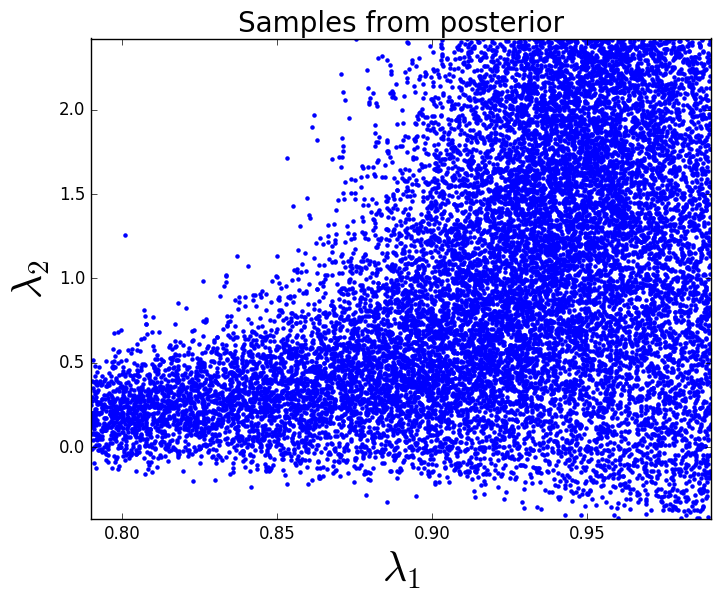}}
	  \scalebox{0.3}{\includegraphics{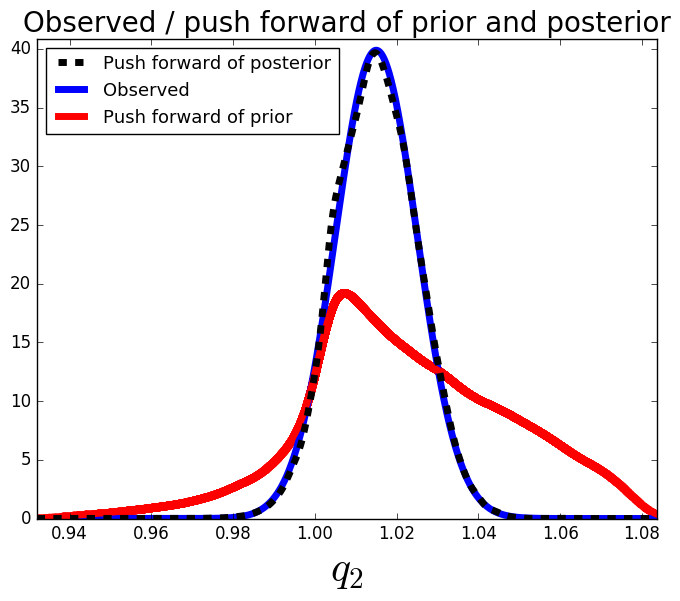}}
    \caption{Approximation of the posterior density obtained using $Q_2$ (left) which gives $I_{Q_2}(\obs)\approx 0.466$, a set of samples from the posterior (middle), and a comparison of the observed density on $Q_2$ with the push-forward densities of the prior and the posterior (right).}\label{fig:2ndqoi}
  \end{center}
\end{figure}

Next, we consider a different QoI to use in the inverse problem, and compare the support of its posterior to the one we just observed.
Specifically consider,
\begin{equation*}
  Q_2(\lambda) = x_1.
\end{equation*}
We assume the observed density on $Q_2$ is a truncated normal distribution with mean 1.015 and standard deviation of 0.01.
%Notice the standard deviation for $Q_2$ is the same as for $Q_1$, implying our experimental uncertainties in $Q_2$ are the same as $Q_1$.
We approximate the push-forward density and the posterior using the same 40,000 samples and again generate a set of samples from the posterior and propagate these samples through the model to approximate the push-forward of the posterior, see Figure~\ref{fig:2ndqoi}.

Although both $Q_1$ and $Q_2$ have the same standard deviation in their observed densities, clearly the two QoI produce very different posterior densities.  The posterior corresponding to data from $Q_2$ has a much larger region of support within the parameter space compared to that of the posterior corresponding to $Q_1$.  This is quantified with the information gain from this posterior $I_{Q_2}(\obs)\approx 0.466$.  Given these two maps, $Q_1$ and $Q_2$, and the specified observed data on each of these data spaces, the data $Q_1$ is more {\it informative} of the parameters than the data $Q_2$.

\begin{figure}[htbp]
  \begin{center}
      \scalebox{0.3}{\includegraphics{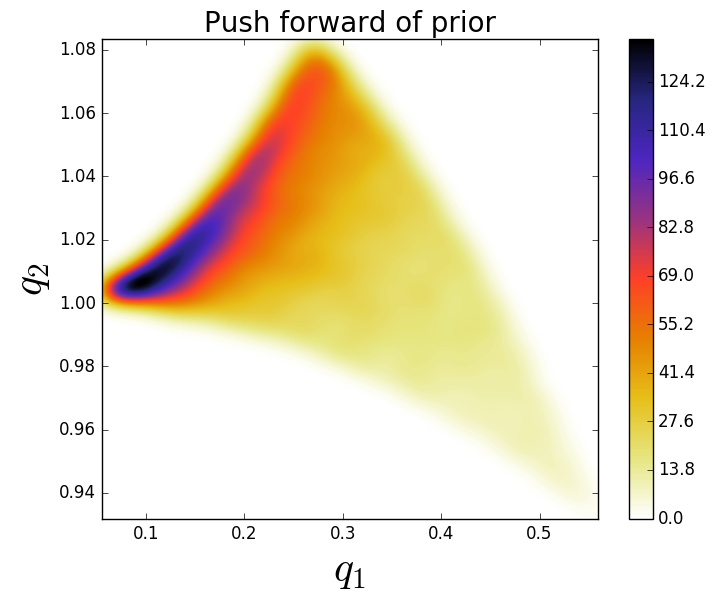}}
      \scalebox{0.3}{\includegraphics{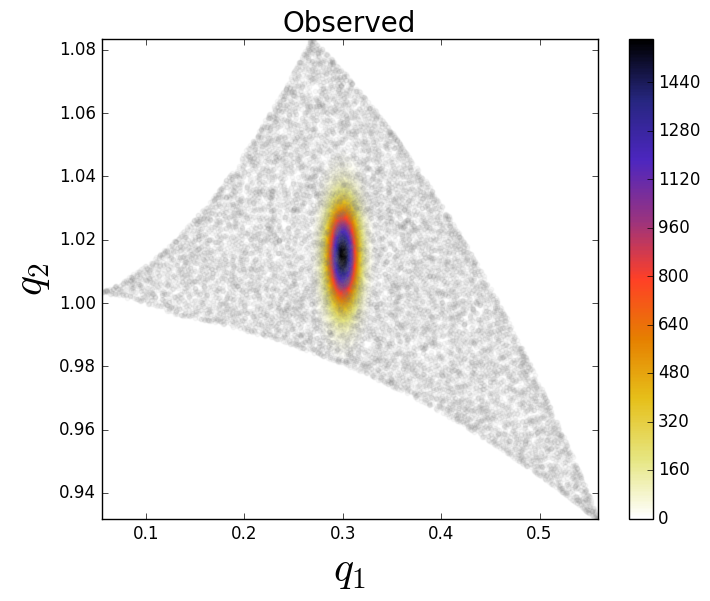}}
      \scalebox{0.3}{\includegraphics{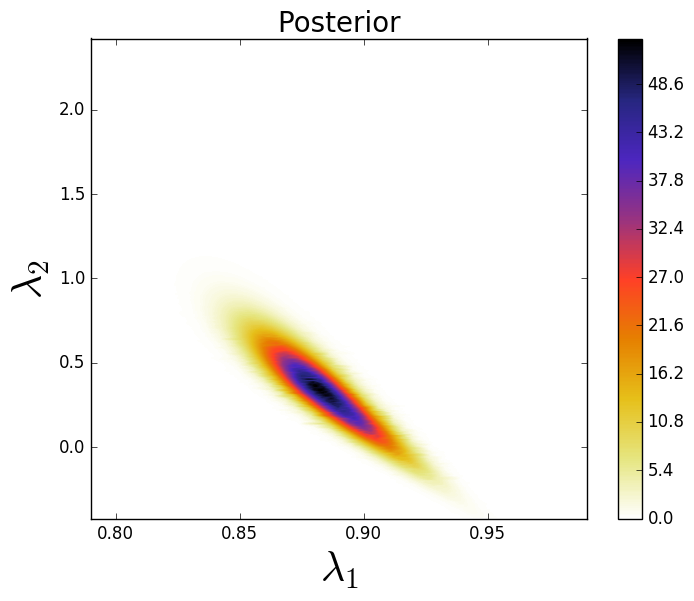}}
    \caption{ The approximation of the push-forward of the prior (left),
the exact observed density on $(Q_1,Q_2)$ (middle),
the approximation of the posterior density using both $Q_1$ and $Q_2$ (right) which gives $I_{Q}(\obs)\approx 2.98$.}\label{fig:bothqoi}
  \end{center}
\end{figure}

Next, we consider using the data from both $Q_1$ and $Q_2$, $Q:\pspace\to(Q_1,Q_2)$, with the same means and standard deviations as specified above.  Again, we approximate the push-forward density and the posterior using the same 40,000 samples, see Figure~\ref{fig:bothqoi}.
With the information from both $Q_1$ and $Q_2$ we see a substantial decrease in the support of the posterior density.  Intuitively, the support of the posterior using both $Q_1$ and $Q_2$ is the support of the posterior using $Q_1$ intersected with the support of the posterior using $Q_2$.  This is quantified in the information gain of this posterior $I_{Q}(\obs)\approx 2.98$.

In the scenario in which we can afford to gather data on both $Q_1$ and $Q_2$, we benefit greatly in terms of reducing the uncertainties on the model input parameters.  However, suppose we could only afford to gather one of these QoI in the field.  Based on the information gain from each posterior, $Q_1$ is more informative about the parameters than $Q_2$.  However, consider a scenario in which the observed data has different means in both $Q_1$ and $Q_2$.  Due to the nonlinearities of the maps, it is not necessarily true that $Q_1$ is still more informative than $Q_2$.  If we do not know the mean of the data for either $Q_1$ or $Q_2$, then we want to determine which of these QoI we {\it expect} to produce the most informative posterior.

%%%%%%%%%%%%%%%%%%%%%%%%%%%%%%%%%%%%%%%%%%%%%%%
\subsection{Expected information gain} \label{sec:expected}
Optimal experimental design must select a design before experimental
data becomes available. In the absence of data we use the simulation model to
quantify the {\it expected} information gain of a given experimental design.
Let $\mathcal{O}$ denote the space of densities over $\D$.  We want to define the expected information gain as some kind of average over this density space in a meaningful way.  However, this is far too general of a space to use to define the expected information gain.  This space includes densities that are unlikely to be observed in reality.  Therefore, we restrict $\mathcal{O}$ to be a space more representative of densities that may be observed in reality.

With no experimental data available to specify an observed density on a single QoI, we assume the density is a truncated Gaussian with a standard deviation determined by some estimate of the measurement instrument error.
%\footnote{We can also consider the standard deviation of the observed data to be uncertain, in which case we average over some interval of possible values for $\sigma$.  However, in this work we assume the uncertainties in the data to be known.}.
With Gaussians of (possibly) varying standard deviations specified for each QoI, this defines the shape of the observed densities we consider.  We let $\mathcal{O}_{\D}$ denote the space of all densities of this shape centered in $\D=Q(\pspace)$,
\begin{equation}\label{eq:od}
\mathcal{O}_{\D} = \Bigg\{\hat{N}(q,\sigma^2) : q\in\D\Bigg\},
\end{equation}
where $\hat{N}(q, \sigma^2)$ is a truncated Gaussian function with mean $q$ and standard deviation $\sigma$.  More details of this definition of $\mathcal{O}_{\D}$ are addressed in Section~\ref{sec:infeasible}.
We can easily generalize our description of $\mathcal{O}$.
For example, we could also consider the standard deviation of the observed data to be uncertain, in which case we would also average over some interval of possible values for $\sigma$.
However, in this work we only vary the center of the Gaussian densities.
\vskip 5pt
\begin{remark}
  We can restrict $\mathcal{O}$ in other ways as well. For example, if we expect the uncertainty in each QoI to be described by a uniform density, then we define the restriction on $\mathcal{O}$ accordingly.
This choice of characterization of the observed density space is largely dependent on the application.
The only limitation is that we require the measure specified on the observed density space to be defined in terms of the push-forward measure, $\pfpriormeas$, as described below.  In Section~\ref{sec:uncertain_source_location} we describe one approach for defining a restricted observed density space where the observed density of each QoI has a Gaussian profile and the standard deviations are functions of the magnitudes of each QoI.
\end{remark}
\vskip 5pt
The restriction of possible $\obs$ to this specific space of densities allows us to represent each density uniquely with a single point $q\in\D$.  Based on our prior knowledge of the parameters and the sensitivities of the map $Q$, the model informs us that some data are more likely to be observed than other data, this is seen in the plot of $\push$ in Figure~\ref{fig:bothqoi} (upper left).  This implies we do not want to average over $\D$ with respect to $\mu$ or $\muD$, but rather with respect to the push-forward of the prior on $D$, $\pfpriormeas$.
This respects the prior knowledge of the parameters and the sensitivity information provided by the model.  We define the {\it expected information gain}, denoted $E(I_Q)$, as just described,
\begin{equation}\label{eq:expectedkl}
  E(I_Q) := \int_{\D}I_Q(q) \push(q) d\mu = \int_{\D}I_Q(q) d\pfpriormeas.
\end{equation}
From Eq.~\eqref{eq:kl}, $I_Q$ itself is defined in terms of an integral.  The expanded form for $E(I_Q)$ is then an iterated integral,
\begin{equation}\label{eq:expectedkliterated}
  E(I_Q) = \int_{\D}\int_{\pspace}\post(\lambda;q)\log\Bigg({\frac{\post(\lambda;q)}{\prior(\lambda)}}\Bigg) d\muL d\pfpriormeas,
\end{equation}
where we make explicit that $\post$ is a function of the observed density and, by our restriction of the space of observed densities in Eq.~\eqref{eq:od}, therefore a function of $q\in\D$.  We utilize Monte Carlo sampling to approximate the integral in Eq.~\eqref{eq:expectedkl} as described in Algorithm~\ref{alg:expectedkl}.

\begin{algorithm}
  \caption{Approximating the Expected Information Gain of an Experiment}
  \label{alg:expectedkl}
  \begin{enumerate}
    \item Given a set of samples from the prior density: $\lambda^{(i)}$, $i=1,\dots,N$;
    \item Given a set of samples from the push-forward density: $q^{(j)} = Q(\lambda^{(j)})$, $j=1,\dots, N$;
    \item Construct an observed density centered at each $q^{(j)}$.
    %\item Compute the value of the posterior density for each $q^{(j)}$, $\post(\lambda; q^{(j)})$, at $\lambda^{(i)}$, $i=1,\dots,N$;
    \item For $j=1,\dots, M$ approximate $I_Q(q^{(j)})$ using \eqref{eq:kl_dataspace}:
    \begin{equation*}
      I_Q(q^{(j)})\approx \frac{\pmeas(\pspace)}{N}\sum_{i=1}^{N}\frac{\obs(Q(\lambda^{(i)}))}{\push(Q(\lambda^{(i)}))} \log\Bigg(\frac{\obs(Q(\lambda^{(i)}))}{\push(Q(\lambda^{(i)}))}\Bigg)
    \end{equation*}
    \item Compute $E(I_Q) \approx \frac{1}{M}\sum_{j=1}^{M}I_Q(q^{(j)})$;

  \end{enumerate}
\end{algorithm}

%\begin{remark}
  Algorithm~\ref{alg:expectedkl} appears to be a computationally expensive procedure since it requires solving
$M$ stochastic inverse problems  and, as noted in \cite{cbayes}, approximating $\push$ can be expensive.
In \cite{cbayes} and in this paper we use kernel density estimation techniques to approximate $\push$ which does not scale well as the dimension of $\dspace$ increases~\cite{wand1994multivariate}.
On the other hand, for a given experimental design, we only need to compute this approximation once, as each $I_Q$ in Step 4 of Algorithm~\ref{alg:expectedkl} is computed using the same prior and map $Q$ and, therefore, the same $\push$.
In other words, the fact that the consistent Bayes method only requires approximating the push-forward of the
prior implies that this information can be used to approximate posteriors for different observed densities
without requiring additional model evaluations.
This significantly improves the computational efficiency of the consistent Bayesian approach
in the context of OED.
We leverage this computational advantage throughout this paper by considering a discrete set of designs which allows us to compute the push-forward for all of the candidate designs simultaneously.
Utilizing a continuous design space might require computing the push-forward for each iteration of the optimization algorithm, since the designs (locations of the observations) are not known {\em a priori}.
The additional model simulations required to compute the push-forward of the prior at new design points might be intractable if the number of iterations is large, however the need for new simulations may be avoided,
if the new observations can be extracted from archived state-space data. For example, if one stores the finite element solutions of a PDE at all samples of the prior at the first iteration of the design optimization, one can evaluate obsevations at new design locations, which are functionals of this PDE solution, via interpolation using the finite element basis.
%\end{remark}

%%%%%%%%%%%%%%%%%%%%%%%%%%%%%%%%%%%%%%%%%%%%%%%
%%%%%%%%%%%%%%%%%%%%%%%%%%%%%%%%%%%%%%%%%%%%%%%
\subsection{Defining the OED} \label{sec:defining}
We are now in a position for define our OED formulation.
Recall that our experimental design is defined as the set of QoI computed from the model
and we seek the optimal set of QoI to deploy in the field. 
Given a physics based model, prior information on the model
parameters, a space of potential experimental designs, and a generic
description of the uncertainties for each QoI, we define our OED as follows.
\begin{definition}[OED]\label{def:oed}
  Let $\mathcal{Q}$ represent the design space, i.e., the space of all possible experimental designs, and $Q^z\in\mathcal{Q}$ be a specific design.  Then the OED is the $Q^z\in\mathcal{Q}$ that maximizes the expected information gain,
  \begin{equation}\label{eq:qopt}
    Q^{\text{opt}} := \arg \max_{Q^z\in\mathcal{Q}}E(I_{Q^z}).
  \end{equation}
\end{definition}
%In general the design space $\mathcal{Q}$, can either be a continuous
%of discrete space.
As previously mentioned, the focus in this paper is on the utilization of the
consistent Bayesian methodology within the OED framework,
so we do not explore different approaches for solving the optimization
problem given by Definition~\ref{def:oed} and simply find the
optimal design over a discrete set of candidate designs.
%In Se ction~\ref{sec:numerical}, we use a particularly simple
%approach to approximate the solution to the optimization problem for
%several continuous design spaces.

\begin{remark}
Consistent Bayesian inference is potentially well suited to finding OED in continuous
design spaces. Typically OED based upon statistical Bayesian methods
uses Markov Chain Monte Carlo (MCMC) methods to characterize the posterior
distribution. MCMC methods do not provide a functional form for the
posterior but rather only provide samples from the
posterior. Consequently, gradient-free or stochastic gradient-based
optimization methods must be used to find the optimal design. In
contrast consistent Bayesian inference provides a functional form for
the posterior which allows the use of more efficient gradient based optimizers.
Exploring the use of more efficient continuous optimization procedures
will be the subject of future work.
\end{remark}

%either solve the optimization problem (for discrete design spaces) or to approximate the solution (for continuous design spaces).

%%% Local Variables:
%%% mode: latex
%%% TeX-master: "../oed_cbayes"
%%% End:

%%%%%%%%%%%%%%%%%%%%%%%%%%%%%%%%%%%%%%%%%%%%%%%
%%%%%%%%%%%%%%%%%%%%%%%%%%%%%%%%%%%%%%%%%%%%%%%
%%%%%%%%%%%%%%%%%%%%%%%%%%%%%%%%%%%%%%%%%%%%%%%
%%%%%%%%%%%%%%%%%%%%%%%%%%%%%%%%%%%%%%%%%%%%%%%
\section{Infeasible data} \label{sec:infeasible}
The OED procedure proposed in this manuscript is based upon consistent Bayesian inference which requires that the observed measure is absolutely continuous with respect to the push-forward measure induced by the prior and the model (assumption A3).
In other words, any event that we observe with non-zero probability will be predicted using the model and prior with non-zero probability.
During the process of computing $E(I_Q)$, it is possible that we violate this assumption.
Specifically, depending on the mean and variance of the observational density we may encounter $\obs\in\mathcal{O}_{\D}$ such that $\int_{\D}\obs d\mu\ < 1$, i.e., support of $\obs$ extends beyond the range of the map $Q$, see Figure~{\ref{fig:violatea4} (upper right).
In this section we discuss the causes of infeasible data and options for avoiding infeasible data when estimating an optimal experimental design.

\subsection{Infeasible data and consistent Bayesian inference}
When inferring model parameters using consistent Bayesian inference the most common cause for infeasible data is that the model being used to estimate the OED is inadequate. That is, the deviation between the computational model and reality is large enough to prohibit the model from predicting all of the observational data. The deviation between the model prediction and the observational data is often referred to as model structure error and can often be a major source of uncertainty.
This is an issue of most if not all inverse parameter estimation problems~\cite{Kennedy_O_JRSSSB_2001}.
Recently there has been a number of attempts to quantify this error (see e.g., \cite{Sargsyan_NG_IJCK_2015}) however such approaches are beyond the scope of this paper.
In the following we will assume that the model structure error does not prevent the model from predicting all the observational data.
%Model structure error may still be present, however, the approach we present below will be unable to account for this error when inferring model parameters.

\subsection{Infeasible data and OED}
To estimate an approximate OED we must quantify the {\it expected} information gain of a given experimental design (see Section~\ref{sec:expected}).
The expectation is over all possible normal observation densities with mean $q\in\mathcal{D}$ and variance $\sigma$, defined by the space~\eqref{eq:od}.
When the support of $\mathcal{D}$ is bounded these densities may produce infeasible data.
%A bounded measure cannot dominate an unbounded measure.
The effect of this violation increases as $q$ approaches the boundary of $\mathcal{D}$.
%Assuming that our model can predict all possible observational data,
%then the specification of the observation density in ~\eqref{eq:od} may violate (A3).

To remedy this violation of (A3) we must modify the set of observational densities.
In this paper we choose to normalize $\obs$ over $\D$.
We redefine the observed density space $\mathcal{O}_{\D}$ so that (A3) holds for each density in the space,
\begin{equation}\label{eq:osigmanormalized}
\mathcal{O}_{\D} = \Bigg\{\frac{\hat{N}(q,\sigma^2)}{C_q} : q\in\D\Bigg\},
\end{equation}
where $\hat{N}(q, \sigma^2)$ is a truncated Gaussian function with mean $q$ and standard deviation $\sigma$, and $C_q$ is the integral of $\hat{N}(q,\sigma^2)$ over $\D$ with respect to the Lebesgue measure on $\D$,
\begin{equation}
C_q = \int_{\D}\hat{N}(q,\sigma^2)d\mu.
\end{equation}
A similar approach for normalizing Gaussian densities over compact domains was taken in \cite{NME:NME5211}.

%%%%%%%%%%%%%%%%%%%%%%%%%%%%%%%%%%%%%%%%%%%%%%%
\subsection{A nonlinear model with infeasible data} \label{sec:nonlinear}
In this section, we use the nonlinear model introduced in Section~\ref{sec:simple}
to demonstrate that infeasible data can arise from relatively benign assumptions.
Suppose the observed density on $Q_1$ is a truncated normal distribution with mean 0.3 and standard deviation of 0.04.
In this one dimensional data space, this observed density is absolutely continuous with respect to the push-forward of the prior on $Q_1$, see Figure~\ref{fig:violatea4_q1q2} (left).
Next, suppose the observed density on $Q_2$ is a truncated normal distribution with mean 0.982 and standard deviation of 0.01.
Again, in this new one dimensional data space, this observed density is absolutely continuous with respect to the push-forward of the prior on $Q_2$, see Figure~\ref{fig:violatea4_q1q2} (right).
Both of these observe densities are dominated by their corresponding push-forward densities, i.e., the model can reach all of the observed data in each case.

\begin{figure}[htbp]
  \begin{center}
    \scalebox{0.3}{\includegraphics{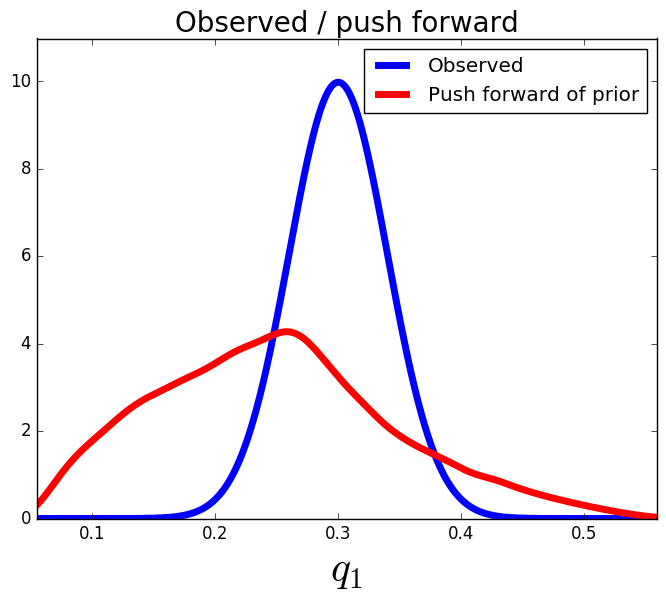}}
      \scalebox{0.3}{\includegraphics{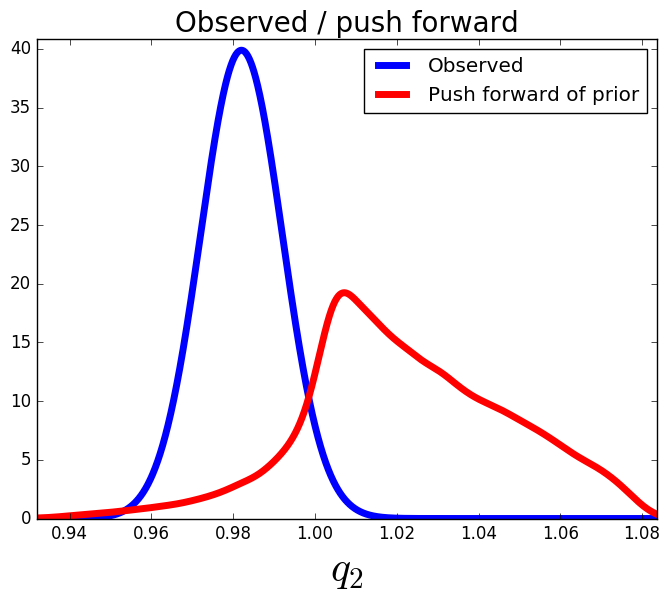}}
    \caption{The push-forward and observed densities on $Q_1$ (left) and the push-forward and observed densities on $Q_2$ (right).  Notice the support of both of the observed densities is contained within the range of the model, i.e., the observed densities are absolutely continuous with respect to their corresponding push-forward densities.}\label{fig:violatea4_q1q2}
  \end{center}
\end{figure}

However, consider the data space defined by {\it both} $Q_1$ and $Q_2$ and the corresponding push-forward and observed densities on this space, see Figure~\ref{fig:violatea4}.
The non-rectangular shape of the combined data space is induced by the nonlinearity in the model and the correlations between $Q_1$ and $Q_2$.
As we see in Figure~\ref{fig:violatea4}, the observed density using the product of the 1-dimensional Gaussian densities is {\it not} absolutely continuous with respect to the push-forward density on $(Q_1, Q_2)$, i.e., the support of $\obs$ extends beyond the support of $\push$.
Referring to Eq.~\eqref{eq:osigmanormalized}, we normalize this observed density over $\D$, see Figure~\ref{fig:violatea4} (right).
Now that the new observed density obeys the assumptions needed, we could solve the stochastic inverse problem as described in Section~\ref{sec:consitent}.

\begin{figure}[htbp]
  \begin{center}
    \scalebox{0.3}{\includegraphics{plots/nonlinear_cbayespaper_pushforward_bothqoi_200grid}}
    \scalebox{0.3}{\includegraphics{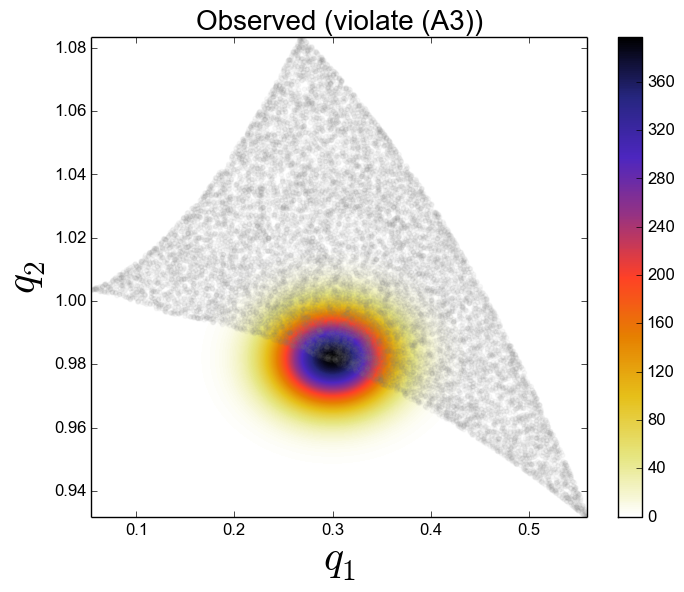}}
	  \scalebox{0.3}{\includegraphics{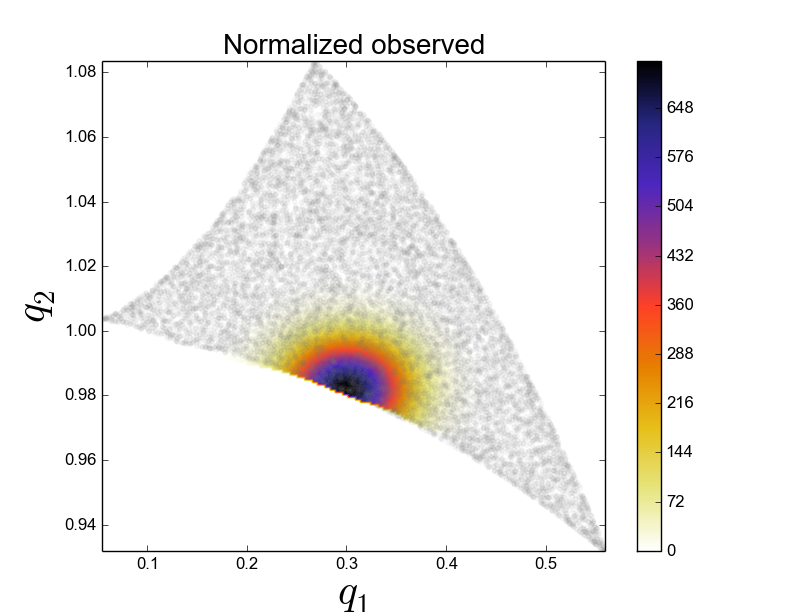}}
\caption{The push-forward of the prior for the map $Q:\pspace\to(Q_1,Q_2)$ introduced in Section~\ref{sec:simple} (left), the observed density using the product of the 1-dimensional Gaussian (middle) which extends beyond the range of the map, and the normalized observed density that does not extend beyond the range of the map (right).}\label{fig:violatea4}
  \end{center}
\end{figure}

%%%%%%%%%%%%%%%%%%%%%%%%%%%%%%%%%%%%%%%%%%%%%%%
\subsection{Computational considerations} \label{sec:computational}
%The OED formulation developed in this paper relies on the consistent Bayesian methodology for solving stochastic inverse problems.
The main computational challenge in the consistent Bayesian approach is the approximation of the push-forward of the prior.
Following \cite{cbayes}, we use Monte Carlo sampling for the forward propagation of uncertainty.
While the rate of convergence is independent of the number of parameters (dimension of $\pspace$), the accuracy in the statistics for the QoI may be relatively poor unless a large number of samples can be taken.
Alternative approaches based on surrogate models can significantly improve the accuracy, but are generally limited to small number of parameters.
We also employ kernel density estimation techniques to construct a non-parametric approximation of the push-forward density, but it is well-known that these techniques do not scale well with the number of observations (dimension of $\dspace$)~\cite{wand1994multivariate}.

Next, we address the computational issue of normalizing $\hat{N}(q,\sigma^2)$, i.e., $\obs$, over $\D$.
From the plot of $\push$ in Figure~\ref{fig:violatea4} (left) it is clear the data space may be a complex region.
Normalizing $\obs$, as in Figure~\ref{fig:violatea4} (right), over $\D$ would be computationally expensive.
Fortunately, the consistent Bayesian approach provides a means to avoid this expense.
Note that from Eq.~\eqref{eq:setbayes} we have,
\begin{equation}\label{eq:lambdabayes}
  \postmeas(\pspace) = \priormeas(\pspace)\frac{\obsmeas(Q(\pspace))}{\pfpriormeas(Q(\pspace))},
\end{equation}
where $\priormeas(\pspace)=\pfpriormeas(Q(\pspace))=1$ which implies,
\begin{equation}\label{eq:lambdabayes1}
  \postmeas(\pspace) = \obsmeas(Q(\pspace)).
\end{equation}
Therefore, normalizing $\obs$ over $\D$ is equivalent to solving the inverse problem and then normalizing $\tildepost$ (where we use the tilde over $\pi$ to indicate this function does not integrate to 1 because we have violated (A3)) over $\pspace$.  Although $\pspace$ may not always be a generalized rectangle, (A1) implies we have a clear definition of $\pspace$ and therefore can efficiently integrate $\tildepost$ over $\pspace$ and then normalize $\tildepost$ by
\begin{equation}\label{eq:normalize}
  \post = \frac{\tildepost}{\int_{\pspace}\tildepost d\muL}.
\end{equation}
In fact, this normalization factor can be estimated without additional model evaluations and without using the values of the prior or the posterior, which may not be usable in high-dimensional spaces.
We observe that
\[\postmeas(\pspace) = \int_\pspace \postdens \ d\muL = \int_\pspace \frac{\obsdens(Q(\lambda))}{\pfpriordens(Q(\lambda))} \ d\priormeas.\]
Thus, we can use the values of $\obsdens$ and $\pfpriordens$ computed for the
samples generated from the prior, which were used to estimate the push-forward of prior, to integrate $\obsdens/\pfpriordens$ with respect to the prior.

%%%%%%%%%%%%%%%%%%%%%%%%%%%%%%%%%%%%%%%%%%%%%%%
%%%%%%%%%%%%%%%%%%%%%%%%%%%%%%%%%%%%%%%%%%%%%%%
%%%%%%%%%%%%%%%%%%%%%%%%%%%%%%%%%%%%%%%%%%%%%%%
%%%%%%%%%%%%%%%%%%%%%%%%%%%%%%%%%%%%%%%%%%%%%%%
\section{Numerical examples} \label{sec:numerical}
In this section we consider several models of physical systems.
%The first example, a linear model, is a bit abstract.
%We assume we have some model of a physical system where the QoI are linear functions of the parameters.
First, we consider a stationary convection-diffusion model with a single uncertain parameter controlling the
magnitude of the source term.
Next, we consider a transient transport model with a two dimensional parameter space determining the location
of the source of a contaminant.
Then, we consider a inclusion problem in computational mechanics where two uncertain parameters control the
shape of the inclusion.
Finally, we consider a high-dimensional example of single-phase incompressible flow in porous media where the uncertain permeability field is given by a Karhunen-Loeve expansion~\cite{zhang2004efficient}.

In each example, we have a parameter space $\pspace$, a set of possible QoI, and a specified number of QoI we can afford to gather during the experiment.
This in turn defines a design space $\Q$ and we let $Q^z\in\Q$ represent a single experimental design and $\D^z=Q^z(\pspace)$ the corresponding data space.
For each experimental design, we let $\sigma^z$ represent the standard deviations defined by the uncertainties in each QoI that compose $Q^z$ and $\mathcal{O}_{\D^z}$ represent the observed density space.
%, $\obsz\in\mathcal{O}_{\D^z}$ represent an arbitrary observed density, and $\postz$ represent the corresponding posterior density.

%The first example has a discrete design space and therefore we simply select the OED.
All of these examples have continuous design spaces, so we approximate the OED
by selecting the OED from a large set of candidate designs.
This approach was chosen because it is much more efficient
to perform the forward propagation of uncertainty using random sampling only once
and to compute all of the candidate measurements for each of these random samples.
Alternatively, one could pursue a continuous optimization formulation which would require
a full forward propagation of uncertainty for each new design.
As mentioned in Section~\ref{sec:defining}, one could limit the number of designs using a gradient-based or
Newton-based optimization approach, but this is beyond the scope of this paper.

%%%%%%%%%%%%%%%%%%%%%%%%%%%%%%%%%%%%%%%%%%%%%%%
\subsection{Stationary convection-diffusion: uncertain source amplitude} \label{sec:source_amp}
In this section we consider a convection-diffusion problem with a single uncertain parameter
controlling the magnitude of a source term.
This example serves to demonstrate that the OED formulation gives intuitive results
for simple problems.

%%%%%%%%%%%%%%%%%%%%%%%%%%%%%%%%%%%%%%%%%%%%%%%
\subsubsection{Problem setup} \label{sec:problem_source_amp}
Consider a stationary convection diffusion model on a square domain:
\begin{equation}\label{eq:conv_diff}
 \begin{cases}
  -D\nabla^2 u + \nabla \cdot (v u) = S, & x\in\Omega,\\
  \nabla u \cdot \mathbf{n} = 0, & x\in\Gamma_N \subset \partial \Omega,\\
  u = 0, & x\in\Gamma_D \subset \partial \Omega,
\end{cases}
\end{equation}
with
\begin{equation*}
  S(x) = A\exp\Big(-\frac{||x_{src}-x||^2}{2h^2}\Big)
\end{equation*}
where $\Omega = [0,1]^2$, $u$ is the concentration field, the diffusion coefficient $D=0.01$, the convection vector $v=[1,1]$, and $S$ is a Gaussian source with the following parameters: $x_{src}$ is the location, $A$ is the amplitude, $h$ is the width.
We impose homogeneous Neumann boundary conditions on $\Gamma_N$ (right and top boundaries) and homogeneous Dirichlet conditions on $\Gamma_D$ (left and bottom boundaries).
For this problem, we choose $x_{src} = [0.5,0.5]$, and $h=0.05$.  We let $A$ be uncertain within $[50, 150]$, thus the parameter space for this problem is $\pspace=[50, 150]$.  Hence, our goal is to gather some limited amount of data that provides the best information about the amplitude of the source, i.e., reduces our uncertainty in $A$.
To approximate solutions to the PDE in Eq.~\ref{eq:conv_diff} given a source amplitude $A$,
we use a finite element discretization with continuous piecewise bilinear basis functions
defined on a uniform ($25\times 25$) spatial grid.

%%%%%%%%%%%%%%%%%%%%%%%%%%%%%%%%%%%%%%%%%%%%%%%
\subsubsection{Results} \label{sec:results_amp}
We assume that we have limited resources for gathering experimental data, specifically, we can only afford to place one sensor in the domain to gather a single concentration measurement.  Our goal is to place this single sensor in $\Omega$ to maximize the expected information gained about the amplitude of the source.  We discretize $\Omega$ using 2,000 uniform random points which produces a design space with 2,000 possible experimental designs.  For this problem, we let the uncertainty in each QoI be described by a truncated Gaussian profile with a fixed standard deviation of 0.1.  This produces observed density spaces, $\mathcal{O}_{\D^z}$, as described in Eq.~\ref{eq:osigmanormalized}.

We generate 5,000 uniform samples from the prior and simulate measurements of each QoI for each of these 5,000 samples.  We consider approximate solutions to the OED problem using subsets of the 5,000 samples of size 50, 200, 1,000 and 5,000.
For each experimental design, we calculate $E(I_{Q^z})$ using Algorithm~\ref{alg:expectedkl} and plot $E(I_{Q^z})$ as a function of the discretized design space in Figure~\ref{fig:source_amp}.
Notice the expected information gain is greatest near the center of the domain (near the location of the source) and in the direction of the convection vector away from the source.
This result matches intuition, as we expect data gathered in regions of the domain that exhibit sensitivity to the parameters to produce high expected information gains.

We note that, for this example, a sufficiently accurate approximation to the design space and the OED is obtained using only 50 samples corresponding to 50 model evaluations.
In Table~\ref{tab:source_amp} we show the top 5 experimental designs (computed using the full set of 5,000 samples) and corresponding $E(I_{Q^z})$ for each set of samples.
\begin{figure}
    \centering
    \scalebox{0.4}{\includegraphics{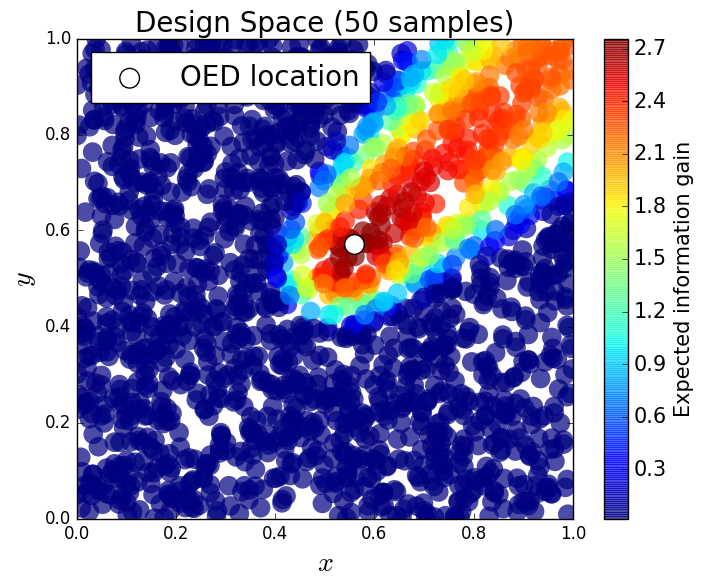}}
    	\scalebox{0.4}{\includegraphics{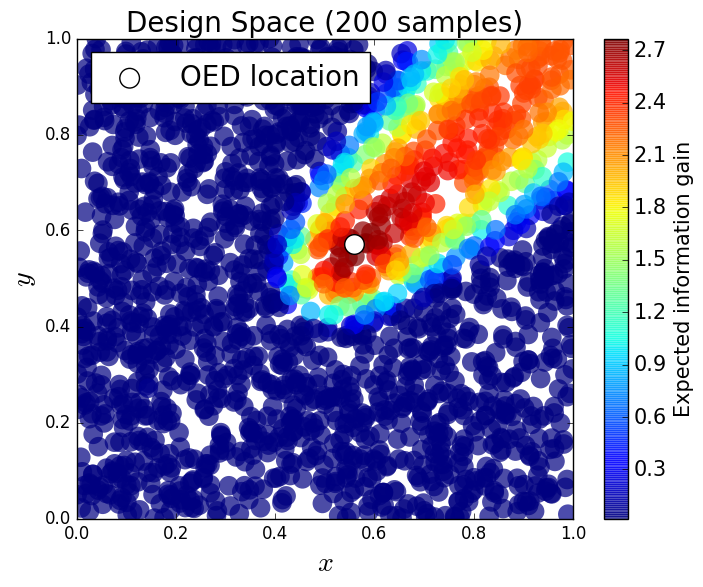}}\\
    \scalebox{0.4}{\includegraphics{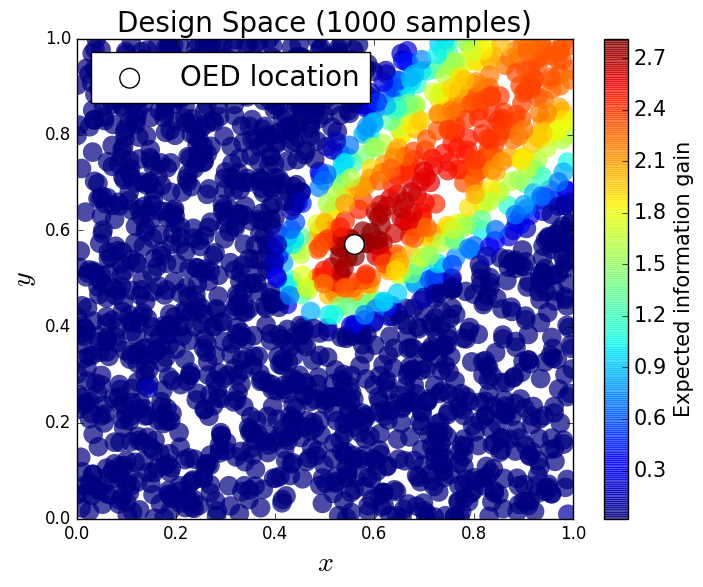}}
    	\scalebox{0.4}{\includegraphics{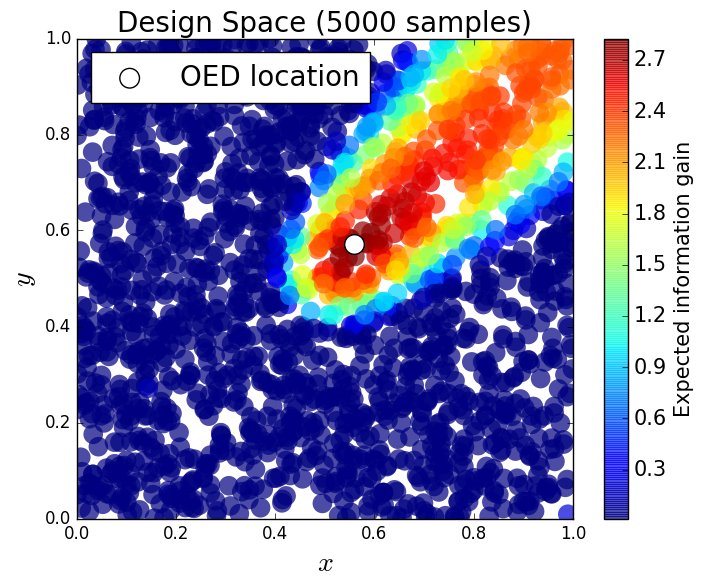}}
\caption{The expected information gain over the design space (which is $\Omega$ in this example) approximated using 50, 200, 1,000 and 5,000 samples from the prior.  Notice the higher values in the center of the domain and towards the top right (in the direction of the convection vector from the location of the source), this is consistent with our intuition.  Moreover, notice the small changes in the design space as we increase the number of samples from 50 to 5,000.  This suggests we compute accurate approximations to the design space using as few as 50 model evaluations.}
\label{fig:source_amp}
\end{figure}

\begin{table}
\centering
\begin{tabular}[ht]{ccccc}
\hline
Design Location & 50 & 200 & 1,000 & 5,000 \\ \hline \hline
 $(0.558, 0.571)$ & 2.758 & 2.767 & 2.815 & 2.826 \\
 $(0.561, 0.546)$ & 2.752 & 2.762 & 2.809 & 2.820 \\
 $(0.582, 0.574)$ & 2.729 & 2.736 & 2.782 & 2.793 \\
 $(0.549, 0.570)$ & 2.728 & 2.735 & 2.781 & 2.792 \\
 $(0.593, 0.596)$ & 2.726 & 2.733 & 2.779 & 2.790 \\ \hline
\end{tabular}
\caption{The top 5 experimental designs chosen using the full set of 5,000 samples.  For each of these designs, we compute $E(I_{Q^z})$ for 50, 200, 1,000 and 5,000 samples.  Notice the change in $E(I_{Q^z})$ for a given design decreases as we increase to 5,000 samples.}
\label{tab:source_amp}
\end{table}

%\begin{table}
%\centering
%\begin{tabular}[ht]{ccc}
%\hline
%Design location & $E(I_{Q^{(z)}})$ \\ \hline \hline
%$(0.56,  0.57)$ & 0.9381 \\
%$(0.56,  0.55)$ & 0.9374 \\
%$(0.58 ,  0.57)$ & 0.9341 \\
%$(0.55 ,  0.57)$ & 0.9340 \\
%$(0.59,  0.60)$ & 0.9338 \\
%$(0.59,  0.57)$ & 0.9337 \\ \hline
%\end{tabular}
%\caption{The expected information gain for each of the top 6 experimental designs.}
%\label{tab:source_amp}
%\end{table}

%%%%%%%%%%%%%%%%%%%%%%%%%%%%%%%%%%%%%%%%%%%%%%%
\subsection{Time dependent diffusion: uncertain source location} \label{sec:uncertain_source_location}
In this section, we compare results from a statistical Bayesian
formulation of OED to the formulation described in this paper.
Specifically, we consider the model in \cite{huan} where the
author uses a classical Bayesian framework for OED to determine the optimal placement of a single sensor that maximizes the expected information about the location of a contaminant source.

%%%%%%%%%%%%%%%%%%%%%%%%%%%%%%%%%%%%%%%%%%%%%%%
\subsubsection{Problem setup} \label{sec:problem_location}
Consider a contaminant transport model on a square domain:
\begin{equation}\label{eq:contaminant}
\begin{cases}
   \frac{\partial u}{\partial t} = \nabla^2 u + S, & x\in\Omega, t>0,\\
  \nabla u \cdot \mathbf{n} = 0, & x\in\partial\Omega, t>0,\\
  u = 0, & x\in\Omega, t=0.
\end{cases}
\end{equation}
with
\begin{equation*}
  S(x) =
  \begin{cases}
    \frac{s}{2\pi h^2}\exp\Big(-\frac{||x_{src}-x||^2}{2h^2}\Big), & \text{if } 0\leq t<\tau, \\
    0, & \text{if } t\geq\tau,
  \end{cases}
\end{equation*}
where $\Omega = [0,1]^2$, $u$ is the space-time concentration field,
we impose homogeneous Neumann boundary conditions along with a zero
initial condition, and $S$ is a Gaussian source with the following
parameters: $x_{src}$ is the location, $s$ is the intensity, $h$ is
the width, and $\tau$ is the shutoff time.

Our goal is to gather some limited amount of data that provides the
best information about the location of the source, i.e., reduces our
uncertainty in $x_{src}$. For this problem, we choose $s=2.0, \,
h=0.05,$ and  $\tau=0.3$ and let $x_{src}$ be uncertain within
$[0,1]^2$ such that $\pspace=[0,1]^2$. To approximate solutions to the PDE in Eq.~\ref{eq:contaminant} given a location of $S$, i.e., a given $x_{src}$, we use a finite element discretization with continuous piecewise bilinear basis functions defined on a uniform ($25\times 25$) spatial grid
and backward Euler time integration with a step size $\Delta t = 0.004$ (100 time steps).

%%%%%%%%%%%%%%%%%%%%%%%%%%%%%%%%%%%%%%%%%%%%%%%
\subsubsection{Results} \label{sec:results_location}
We assume that we have limited resources for gathering experimental data, specifically, we can only afford to place one sensor in the domain and can only gather a single concentration measurement at time $t=0.24$.  Our goal is to place this single sensor in $\Omega$ to maximize the expected information gained about the location of the contaminant source.  For simplicity, we discretize $\Omega$ using an $11\times 11$ regular grid of points which produces a design space with 121 possible experimental designs.  We let the uncertainty in each QoI be described by a Gaussian profile with a standard deviation that is a function of the magnitude of the QoI, i.e.,
\begin{equation}
  \sigma_i = 0.1 + 0.1 |q_i| \hskip 5pt \text{for} \hskip 5pt i=1\dots M,
\end{equation}
where $M$ is the dimension of the data space.  This produces observed density spaces, $\mathcal{O}_{\D^z}$, that consist of truncated Gaussian functions with varying standard deviations,
\begin{equation}
\mathcal{O}_{\D^z} = \Bigg\{\frac{\hat{N}(q,(\sigma(q))^2)}{C_q} : q\in\D^z\Bigg\}.
\end{equation}

We generate 5,000 uniform samples from the prior and simulate measurements of each QoI for each of these 5,000 samples.  We consider approximate solutions to the OED problem using subsets of the 5,000 samples of size 50, 200, 1,000 and 5,000.  For each experimental design, we use this data to calculate $E(I_{Q^z})$ using Algorithm~\ref{alg:expectedkl} and plot $E(I_{Q^z})$ as a function of the discretized design space in Figure~\ref{fig:source_location}.  Notice the expected information gain is greatest near the corners of the domain and smallest near the center, this is consistent with \cite{huan}.  In Table~\ref{tab:source_location} we show the top 5 experimental designs, approximated using the full set of 5,000 samples, and corresponding $E(I_{Q^z})$ for each set of samples.

In Figure~\ref{fig:source_location_posteriors} we consider three different posteriors computed using data from the OED approximated using 5,000 samples, i.e., data gathered by a sensor placed in the bottom left corner of the domain, where each posterior corresponds to a different possible location of the source.  We see varying levels of information gain in these three scenarios, reiterating the point that we choose the OED based on the average of these information gains, $E(I_Q)$.
\begin{figure}
    \centering
    \scalebox{0.4}{\includegraphics{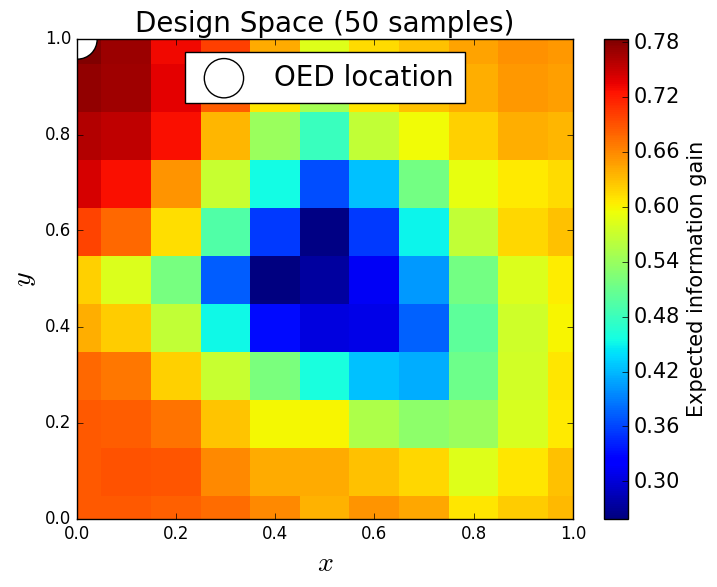}}
    	\scalebox{0.4}{\includegraphics{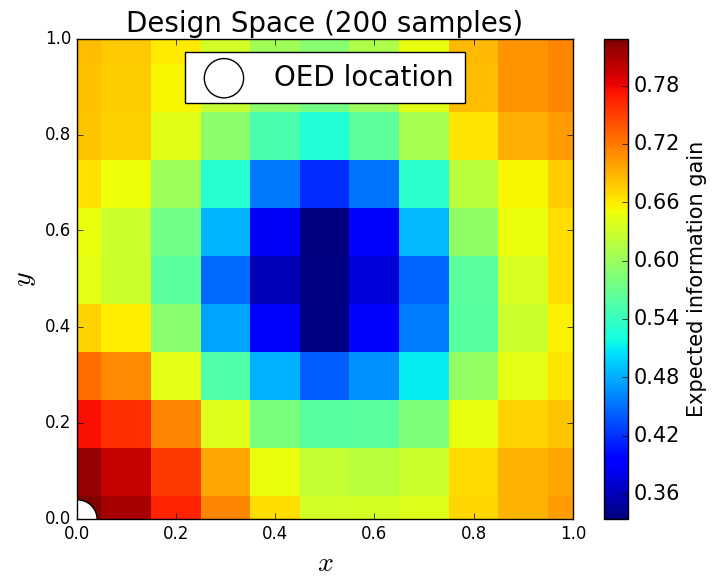}}\\
    \scalebox{0.4}{\includegraphics{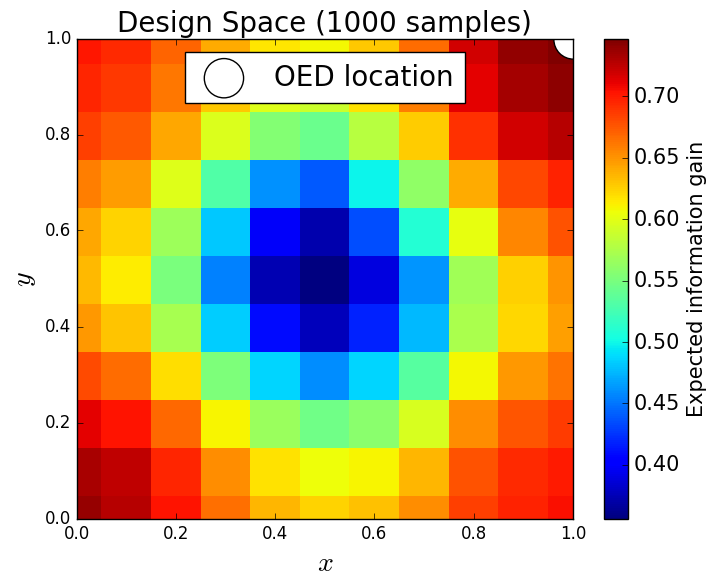}}
    	\scalebox{0.4}{\includegraphics{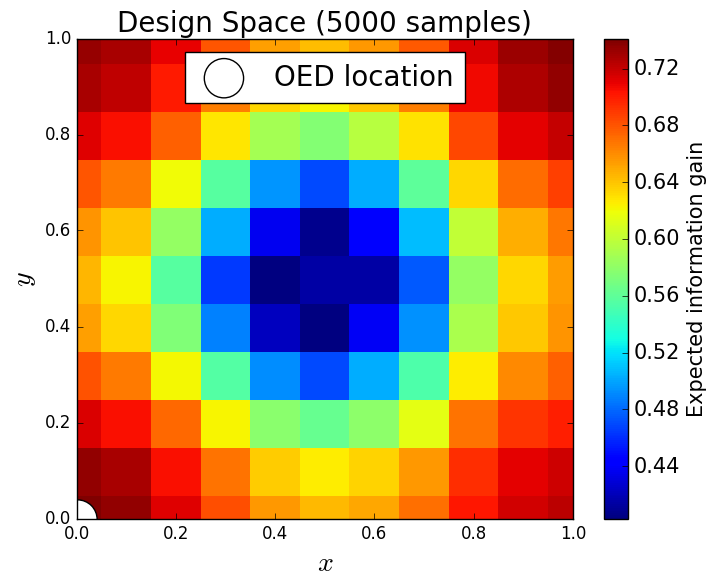}}
\caption{The expected information gain over the design space (which is $\Omega$ in this example) approximated using 50, 200, 1,000 and 5,000 samples from the prior.  Notice the higher values in the corners and the general trend are consistent with \cite{huan}.}
\label{fig:source_location}
\end{figure}

\begin{table}
\centering
\begin{tabular}[ht]{ccccc}
\hline
Design Location & 50 & 200 & 1,000 & 5,000 \\ \hline \hline
$(0, 0)$ & 0.687 & 0.828 & 0.738 & 0.741 \\
$(1, 1)$ & 0.653 & 0.713 & 0.747 & 0.740 \\
$(0, 0.1)$ & 0.687 & 0.817 & 0.733 & 0.736 \\
$(0.1, 0)$ & 0.687 & 0.810 & 0.728 & 0.735 \\
$(1, 0.9)$ & 0.648 & 0.713 & 0.742 & 0.735 \\ \hline
\end{tabular}
\caption{The top 5 experimental designs chosen using the full set of 5,000 samples.  For each of these designs, we compute $E(I_{Q^z})$ for 50, 200, 1,000 and 5,000 samples.  Notice the change in $E(I_{Q^z})$ for a given design decreases as we increase to 5,000 samples.}
\label{tab:source_location}
\end{table}

\begin{figure}[htbp]
  \begin{center}
    \scalebox{0.3}{\includegraphics{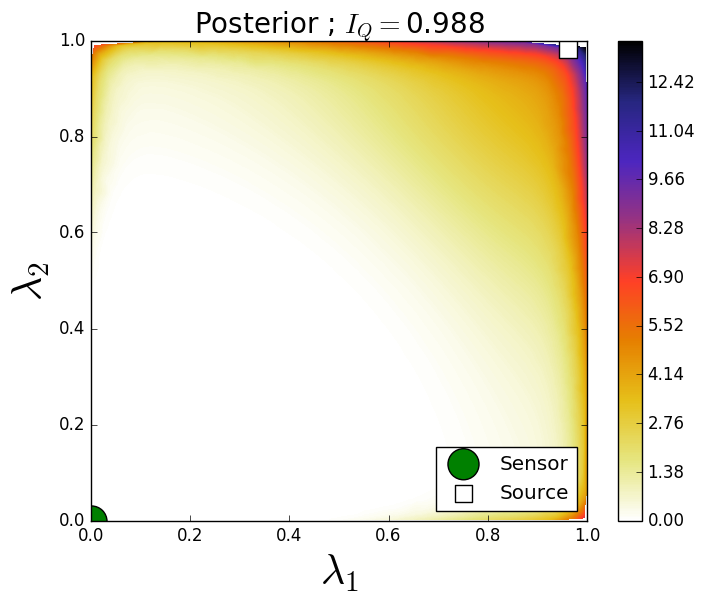}}
      \scalebox{0.3}{\includegraphics{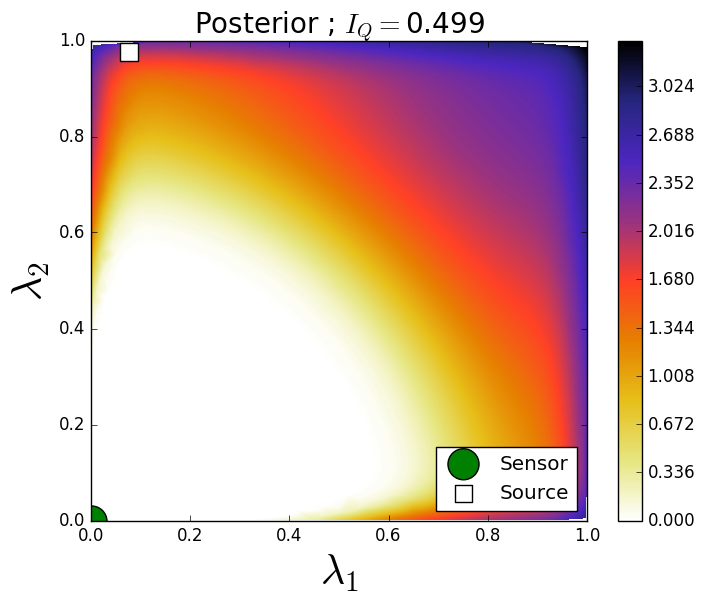}}
	\scalebox{0.3}{\includegraphics{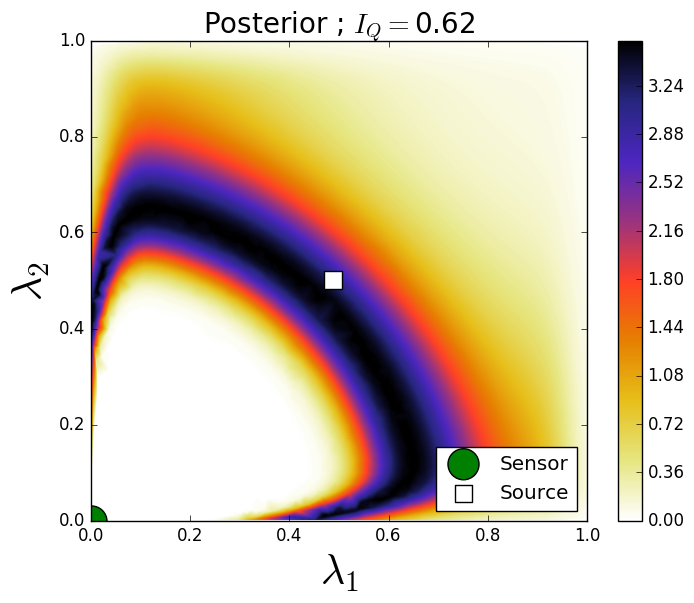}}
    \caption{Posteriors, approximated using 5,000 samples, using the OED for three realizations of the location of the source.  Notice the information gain changes substantially for each posterior, however, this experimental design, placement of the sensor in the bottom left corner, produces the maximum average information gain, $E(I_Q)$, over all possible locations of the source.}\label{fig:source_location_posteriors}
  \end{center}
\end{figure}

\begin{remark}
  Although many of the results in this section seem to match our intuition about which measurement locations should produce high expected information gains, this may not always be the case.  In particular, we have found that our results can depend on our choice of the variance in the the observed densities $\sigma$.  If $\sigma$ is chosen to be large relative to the range of a data space, then the posteriors produced as we average over $\mathcal{O}_{\D}$ are all nearly the same and potentially produce unusually high information gains when the observed densities have substantial support over regions of the data space with very small probability (very small values of the push-forward of the prior).  Another way to think of this is the push-forward densities have high entropy and because $\sigma$ is large $\obs$ is very close to uniform and this produces posterior densities with high information gains.  If $\sigma$ is chosen to be small relative to the range of the data space, i.e., if we expect the experiments to be informative, we do not encounter this issue because we are integrating over $\D$ with respect to the push-forward measure so most of our potential observed data lies in high probability regions of the data space.
%It is worth noting the current limitations of this OED method, however, we do not address this concern in this work.
\end{remark}

%%%%%%%%%%%%%%%%%%%%%%%%%%%%%%%%%%%%%%%%%%%%%%%
\subsection{A Parameterized Inclusion} \label{sec:uncertain_inclusion}
In this section, we consider a simple problem in computational mechanics where the precise boundary
of an inclusion is uncertain.
We parameterize the inclusion and seek to determine the location to place a
sensor that will maximize the information gained regarding the shape of the inclusion.
We use a linear elastic formulation to model the response of the media to surface forces
and measure horizontal stress at each sensor location.
We assume that the material properties (Poisson ratio and Young's modulus) are different
inside the inclusion and that these properties are known a prior.

%%%%%%%%%%%%%%%%%%%%%%%%%%%%%%%%%%%%%%%%%%%%%%%
\subsubsection{Problem setup} \label{sec:problem_inclusion}

Consider a linear elastic plane strain model,
\begin{equation}\label{eq:linelast}
\begin{cases}
-\nabla \cdot \mathbf{\sigma(u)} = \mathbf{0}, & x\in \Omega = [-5,5]\times[0,2],\\
\mathbf{u} = \mathbf{g}, & x\in \Gamma_D = \left\{(x,y)\in \Omega \ | \ x=0\right\},\\
\mathbf{\sigma(u)} \mathbf{n} = \mathbf{t}, & x\in \Gamma_N = \partial \Omega \backslash \Gamma_D,
\end{cases}
\end{equation}
%where $\Omega = [-5,5]\times[0,2]$ and $\partial \Omega = \Gamma_D \cap \Gamma_N$ and
where $\mathbf{\sigma(u)}$ is given by the linear elastic
constitutive relation,
\[\mathbf{\sigma(u)} = \lambda (\nabla \cdot \mathbf{u})\mathbb{I} + \mu (\nabla \mathbf{u} +\nabla \mathbf{u}^T).\]
We express this relation in terms of the Lam\'{e} parameters, $\lambda$ and $\mu$, which are related to
the Poisson ratio, $\nu$, and Young's modulus, $E$, via the following expressions,
\[ \mu = \frac{E}{2(1+\nu)}, \quad \lambda = \frac{E \nu}{(1+\nu)(1-2\nu)}.\]
Now assume that there is an inclusion within the media defined by an ellipse
\[ {\cal I} = \left\{(x,y)\in \Omega \ | \ \frac{1}{\alpha}(x-x_0)^2 + \frac{1}{\beta}(y-y_0)^2 \leq 1 \right\},\]
where $x_0=y_0=0$ and $\alpha$ is uniformly distributed on $[0.5,1]$ and $\beta$ is uniformly distributed on $[0.25,0.5]$.
The material properties are assumed to be known and are given by,
\[\nu = \begin{cases}
0.45, & (x,y)\in {\cal I},\\
0.3, & \text{otherwise},
\end{cases}, \quad
E = \begin{cases}
10.0, & (x,y)\in {\cal I},\\
40.0, & \text{otherwise},
\end{cases}.
\]
These material properties were not chosen to emulate any particular materials, just to demonstrate the proposed
OED formulation.
\begin{figure}
\centering
\scalebox{0.3}{\includegraphics[trim={1cm 4cm 2cm 5cm},clip]{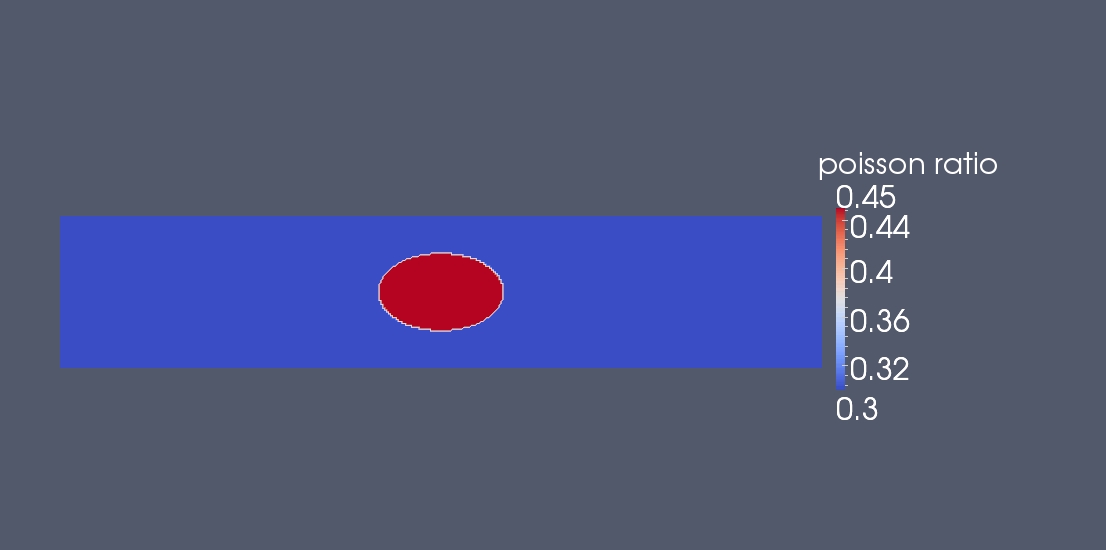}}
\caption{The computational domain and Poisson ratio showing the inclusion for a particular realization of the ellipsoid parameters.}
\label{fig:inclusion_domain}
\end{figure}

Next let us impose homogeneous Dirichlet boundary conditions on the
bottom boundary and stress free boundary conditions on the sides, and
impose a uniform traction in the y-direction along the top boundary
($\mathbf{t}_{\text{top}} = (0,-1)^T$). And finally assume
that we can probe the media and measure the horizontal stress at a given sensor location.
We do not want to puncture the inclusion, so we only consider sensor locations outside the bounds on the inclusion.
Equation~\ref{eq:linelast} was solved using a finite element discretization with piecewise linear basis functions defined on a uniform $400\times 80$ mesh resulting in a system with 64,962 degrees of freedom.
The computational model is implemented using the Trilinos toolkit \cite{Trilinos-Overview} and each realization of the model requires approximately 1 second using 8 processors.

%%%%%%%%%%%%%%%%%%%%%%%%%%%%%%%%%%%%%%%%%%%%%%%
\subsubsection{Results} \label{sec:results_inclusion}

As in previous examples, we assume that we have limited resources for gathering experimental data,
specifically, we can only afford to place one sensor in the domain to gather a single stress measurement.
Our goal is to place this single sensor to maximize the expected information gained about the
shape of the inclusion.
We select 2,000 random sensor locations (outside the inclusion bounds)
which produces a design space with 2,000 possible experimental designs.
For this problem, we let the probability density for the QoI be described by a truncated Gaussian profile with
a fixed standard deviation of 0.001.
We generate 1,000 uniform samples from the prior and compute the horizontal stress at each sensor location
for each of these 1,000 samples.

First, we compare the posterior densities for two sensor locations, $(3.5294,1.3049)$ and $(1.3902,1.2100)$, under the
assumption that we have already gathered data at these sensor locations.
The purpose here is to demonstrate that we obtain different posterior densities and therefore
gain different information from each sensor.
The first sensor is further from the inclusion so we expect that the data from the second sensor will
constrain the posterior more than the data from the first.
In Figures~\ref{fig:inclusion_post_point5} and \ref{fig:inclusion_post_point6}, we plot the samples from the posterior and the corresponding kernel density estimate of the posterior for the first and second sensor locations respectively.
It is clear that measuring the horizontal stress closer to the inclusion increases the information gained from the prior to the posterior.

\begin{figure}[htbp]
  \begin{center}
    \scalebox{0.3}{\includegraphics{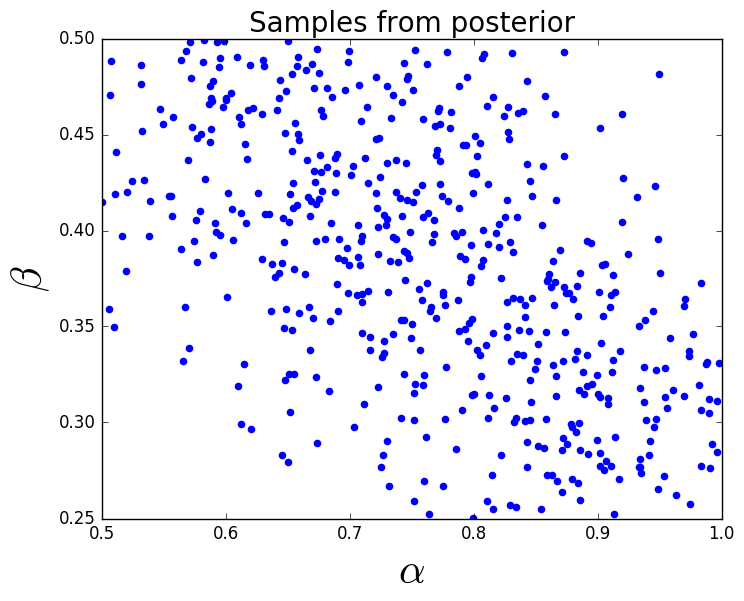}}
      \scalebox{0.3}{\includegraphics{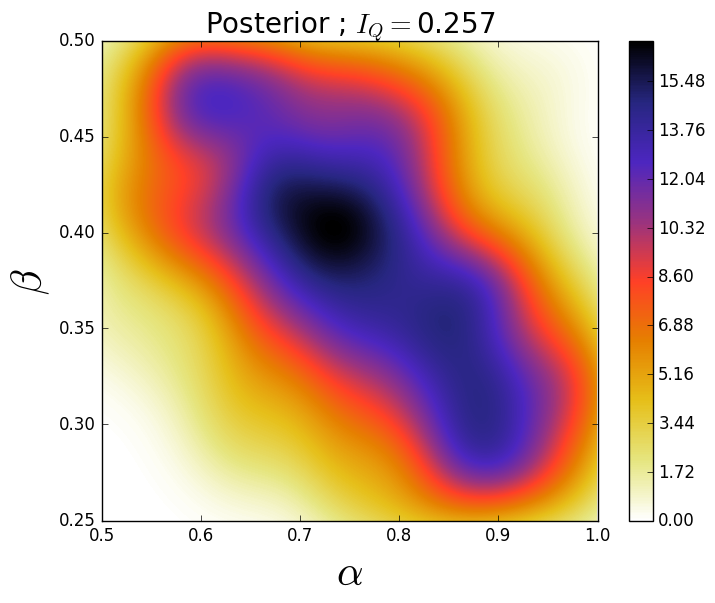}}
    \caption{The set of samples from the posterior and the corresponding kernel density estimate of the posterior for the first sensor location, $(3.5294,1.3049)$.}\label{fig:inclusion_post_point5}
  \end{center}
\end{figure}

\begin{figure}[htbp]
  \begin{center}
    \scalebox{0.3}{\includegraphics{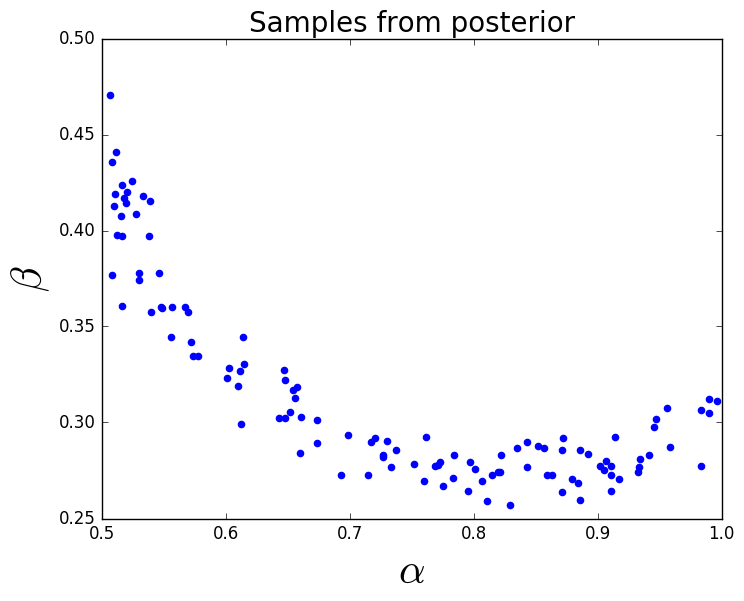}}
      \scalebox{0.3}{\includegraphics{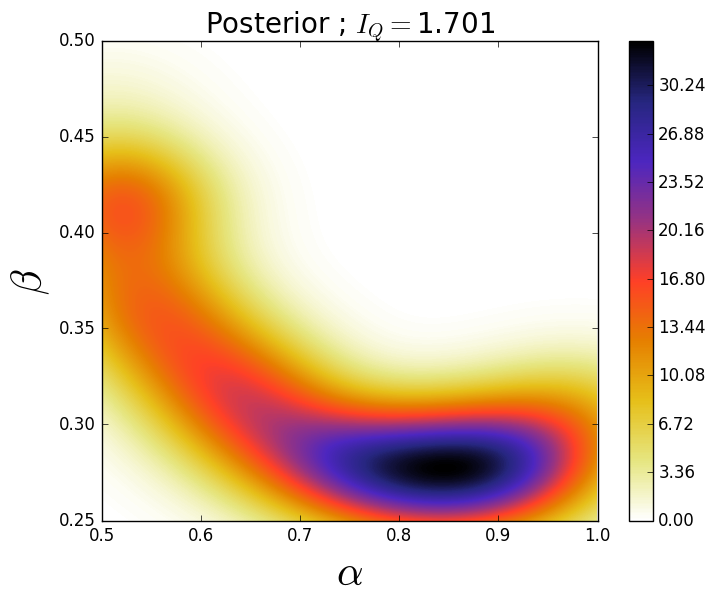}}
    \caption{The set of samples from the posterior and the corresponding kernel density estimate of the posterior for the second sensor location, $(1.3902,1.2100)$.}\label{fig:inclusion_post_point6}
  \end{center}
\end{figure}

We consider approximate solutions to the OED problem using subsets of the 1,000 samples of size 10, 50, 100 and 1,000.  For each experimental design, we use this data to calculate $E(I_{Q^z})$ using Algorithm~\ref{alg:expectedkl} and plot $E(I_{Q^z})$ as a function of the discretized design space in Figure~\ref{fig:inclusion_oed}.  Notice the expected information gain is greatest near the bottom of the domain near the inclusion and is reasonably symmetric around the inclusion.
Also note that the expected information gain is relatively large in the bottom corners of the domain.
This is due to the choice of boundary conditions for the model which induces a large amount of stress in these corners.  In Table~\ref{tab:source_location} we show the top 5 experimental designs, approximated using the full set of 1,000 samples, and corresponding $E(I_{Q^z})$ for each set of samples.
\begin{figure}
	\centering
	\scalebox{0.4}{\includegraphics{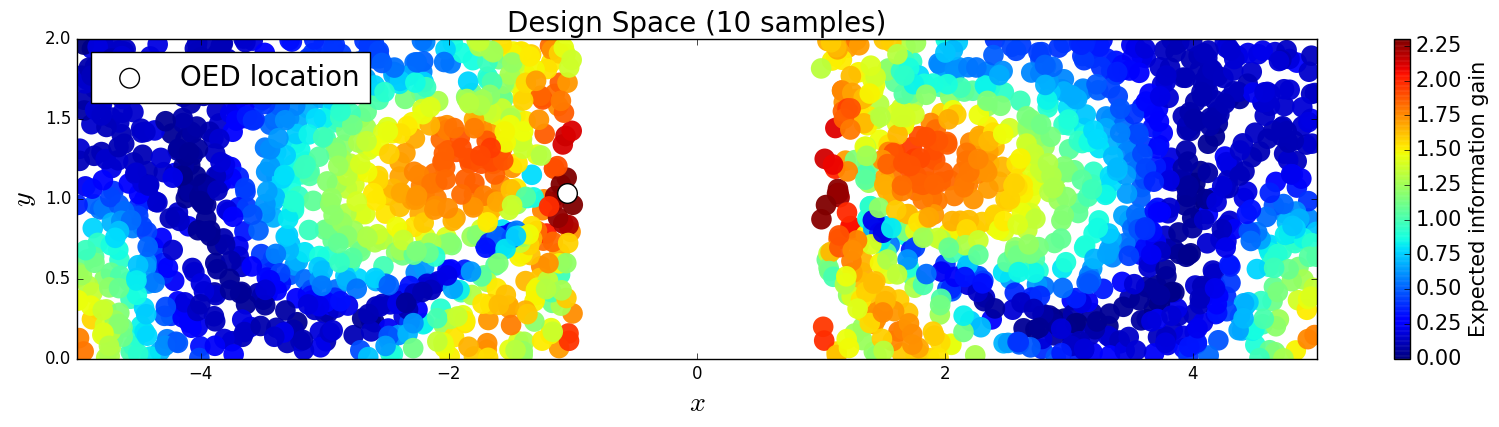}}\\
	\scalebox{0.4}{\includegraphics{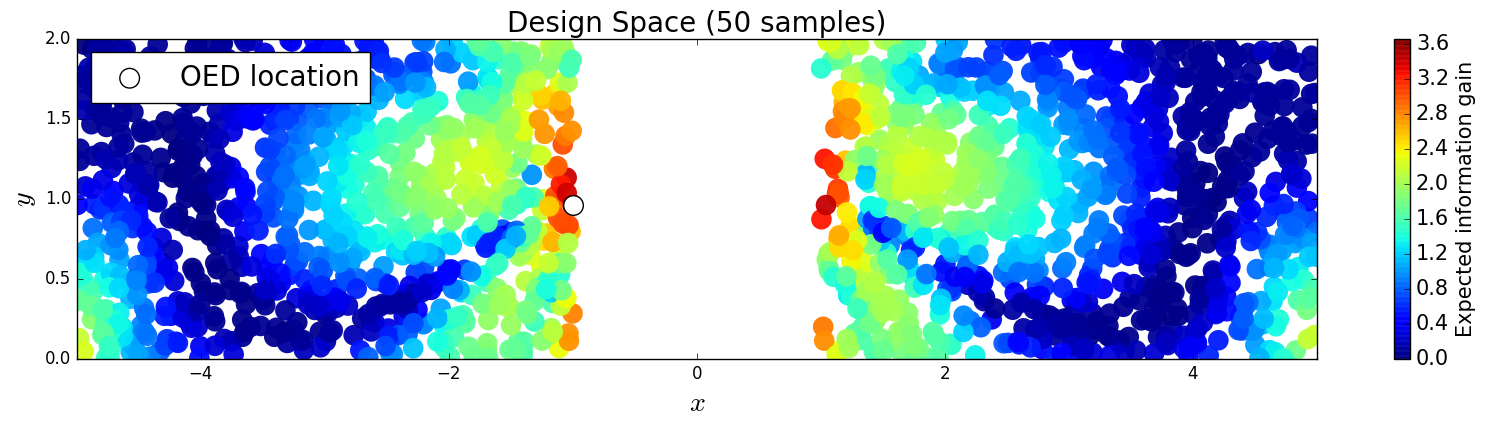}}\\
	\scalebox{0.4}{\includegraphics{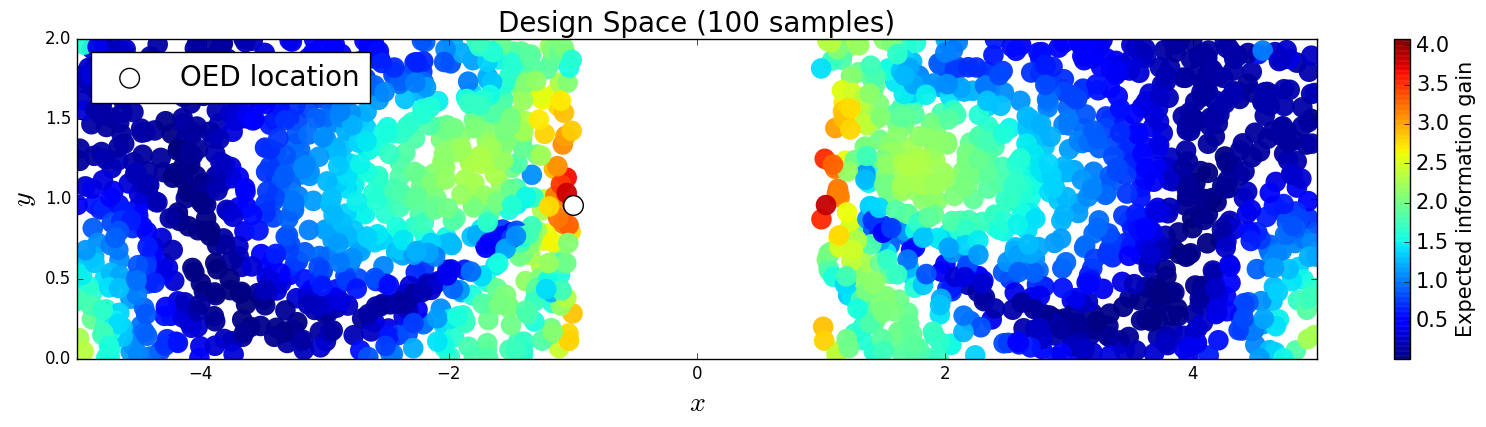}}\\
	\scalebox{0.4}{\includegraphics{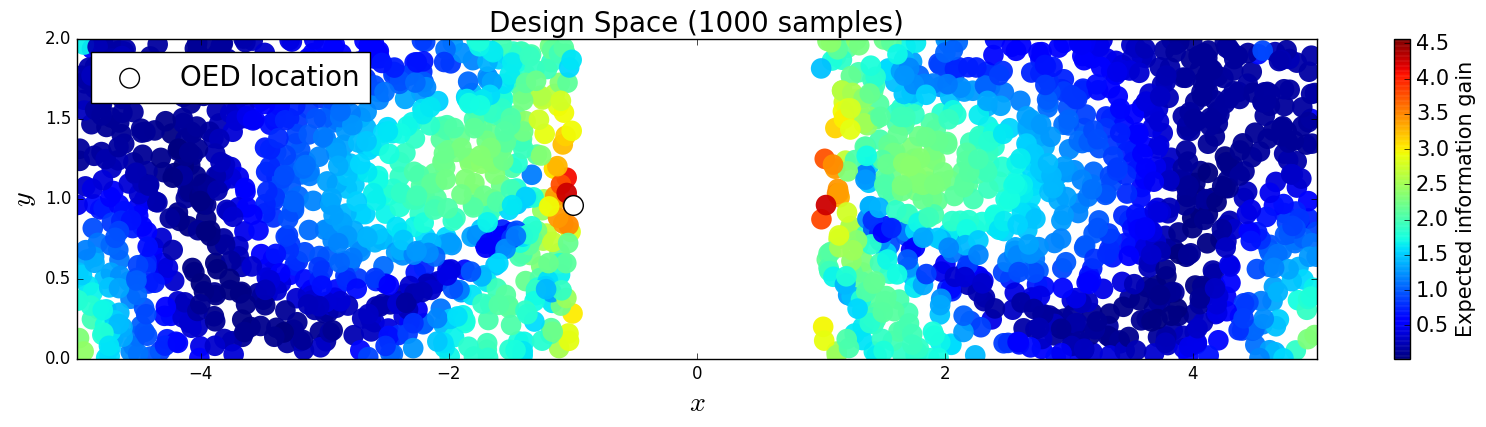}}
	\caption{The expected information gain over the design space (which is $\Omega$ in this example) approximated using 10, 50, 100 and 1,000 samples from the prior.  Notice the higher values near the location of the inclusion.}
	\label{fig:inclusion_oed}
\end{figure}

\begin{table}
\centering
\begin{tabular}[ht]{ccccc}
\hline
Design Location & 10 & 50 & 100 & 1,000 \\ \hline \hline
$(-1.001, 0.961)$ & 2.303 & 3.662 & 4.085 & 4.569 \\
$(-1.050, 1.035)$ & 2.303 & 3.391 & 3.811 & 4.261 \\
$(1.041, 0.959)$ & 2.303 & 3.437 & 3.839 & 4.256 \\
$(-1.050, 1.130)$ & 2.302 & 3.418 & 3.661 & 4.096 \\
$(1.005, 0.870)$ & 2.277 & 3.246 & 3.544 & 3.813 \\\hline
\end{tabular}
\caption{The top 5 experimental designs chosen using the full set of 1,000 samples.  For each of these designs, we compute $E(I_{Q^z})$ for 10, 50, 100 and 1,000 samples.}
\label{tab:inclusion}
\end{table}

%%%%%%%%%%%%%%%%%%%%%%%%%%%%%%%%%%%%%%%%%%%%%%%
\subsection{A Higher-Dimensional Porous Media Example with Uncertain Permeability} \label{sec:high-dimension}
In this section, we consider an example of single-phase incompressible
flow in porous media with a Karhunen-Lo\'eve expansion of the uncertain
permeability field.  The purpose of this example is to demonstrate the OED formulation
on a problem with a high-dimensional parameter space and more than one sensor.

%%%%%%%%%%%%%%%%%%%%%%%%%%%%%%%%%%%%%%%%%%%%%%%
\subsubsection{Problem setup} \label{sec:problem_highd}
Consider a single-phase incompressible flow model:
\begin{equation}\label{eq:porous}
\begin{cases}
-\nabla \cdot (K(\lambda) \nabla p) = 0, & x\in\Omega = (0,1)^2,\\
p = 1, & x=0, \\
p = 0, & x=1, \\
K\nabla p \cdot \mathbf{n} = 0, & y=0 \text{ and } y=1.
\end{cases}
\end{equation}
Here, $p$ is the pressure field and $K$ is the permeability field which we assume is a scalar field given by a Karhunen-Lo\'eve expansion of the log transformation, $Y = \log{K}$, with
\[Y(\lambda) = \overline{Y} + \sum_{i=1}^\infty \xi_i(\lambda)\sqrt{\eta_i}f_i(x,y),\]
where $\overline{Y}$ is the mean field.  We assume the mean removed random media is given by a Gaussian process which implies that the $\xi_i$ are mutually uncorrelated random variables with zero mean and unit variance \cite{ganis2008stochastic,wheeler2011multiscale}.
The eigenvalues, $\eta_i$, and eigenfunctions, $f_i$, are computed numerically using the following covariance function,
\[C_Y({\mathbf x},\overline{{\mathbf x}}) = \sigma_Y^2 \exp \left[ -\frac{(x_1-\overline{x}_1)^2}{2\eta_1} - \frac{(x_2-\overline{x}_2)^2}{2\eta_2}\right],\]
%which was also utilized in \cite{zhang2004efficient,Schwab2006100}.
where $\sigma_Y$ and $\eta_i$ denote the variance and correlation length in the i$^\text{th}$ spatial direction respectively.
We assume a correlation length of $0.01$ in each spatial direction and truncate the expansion at 100 terms.
This choice of truncation is purely for the sake of demonstration.
In practice, the expansion is truncated once a sufficient fraction of the energy in the eigenvalues is retained~\cite{zhang2004efficient,ganis2008stochastic}.
This truncation gives 100 uncorrelated random variables, $\xi_1,\ldots, \xi_100$, with zero mean and unit variance which implies $\pspace = \mathbb{R}^100$.
To approximate solutions to the PDE in Eq.~\ref{eq:porous} we use a finite element discretization with continuous piecewise bilinear basis functions defined on a uniform ($50\times 50$) spatial grid.

%%%%%%%%%%%%%%%%%%%%%%%%%%%%%%%%%%%%%%%%%%%%%%%
\subsubsection{Results} \label{sec:results_highd}
In this section, we present approximate solutions to several different design problems.  We begin with the familiar problem of choosing a single sensor location within the physical domain.  Then, we consider approximating the optimal location of a second sensor {\it given} the location of the first sensor.  In this way, we solve the {\it greedy} OED problem and determine the greedy optimal locations of 1-8 sensors within the physical domain.  We then consider solving the ehaustive OED problem where we limit the sensors to 25 locations and consider determining the optimal location of 5 available sensors.

First, assume that we have limited resources for gathering experimental data, specifically, we can only afford to place one sensor in the domain to gather a single pressure measurement.  Our goal is to place this single sensor in $\Omega$ to maximize the expected information gained about the amplitude of the source.  We discretize $\Omega$ using 1,301 pionts on a grid which produces a design space with 1,301 possible experimental designs.  For this problem, we let the uncertainty in each QoI be described by a truncated Gaussian profile with a fixed standard deviation of 0.01.  This produces observed density spaces, $\mathcal{O}_{\D^z}$, as described in Eq.~\ref{eq:osigmanormalized}.

We generate 10,000 samples from the prior and simulate measurements of each QoI.  We consider approximate solutions to the OED problem using subsets of the 10,000 samples of size 50, 100, 1,000 and 10,000.
For each experimental design, we calculate $E(I_{Q^z})$ using Algorithm~\ref{alg:expectedkl} and plot $E(I_{Q^z})$ as a function of the discretized design space in Figure~\ref{fig:porousmedia}.
Notice the expected information gain is greatest near the top and bottom of the domain away from the left and right edges.  This result matches intuition, as we expect data gathered near the left and right edges to be less informative given the Dirichlet boundary condition imposed on those boundaries.
We note that, for this example, a sufficiently accurate approximation to the design space and the OED is obtained using only 1,000 samples corresponding to 1,000 model evaluations.
In Table~\ref{tab:source_amp} we show the top 5 experimental designs (computed using the full set of 10,000 samples) and corresponding $E(I_{Q^z})$ for each set of samples.

\begin{figure}
    \centering
    \scalebox{0.4}{\includegraphics{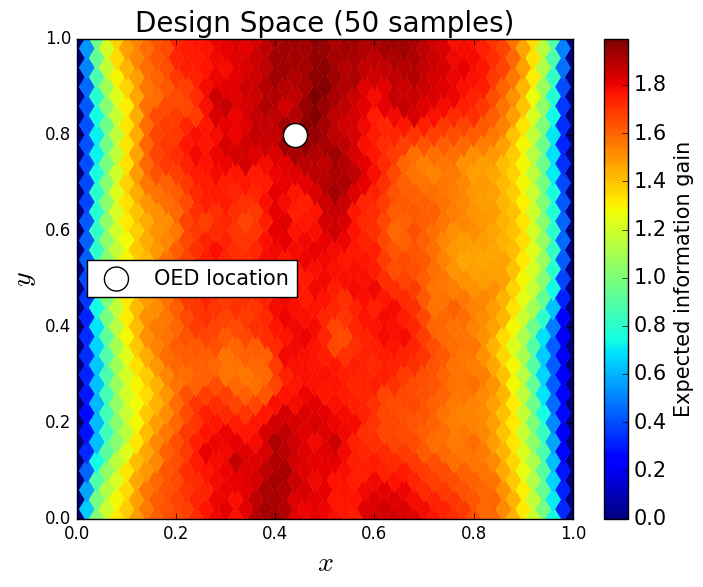}}
    	\scalebox{0.4}{\includegraphics{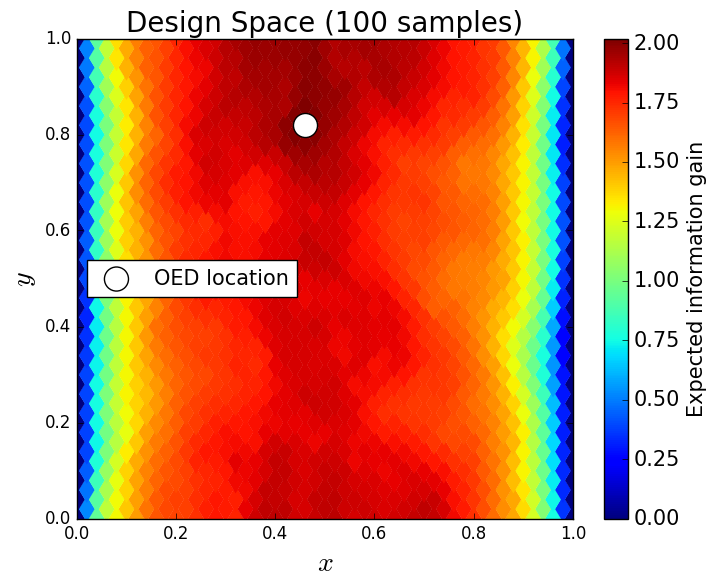}}\\
    \scalebox{0.4}{\includegraphics{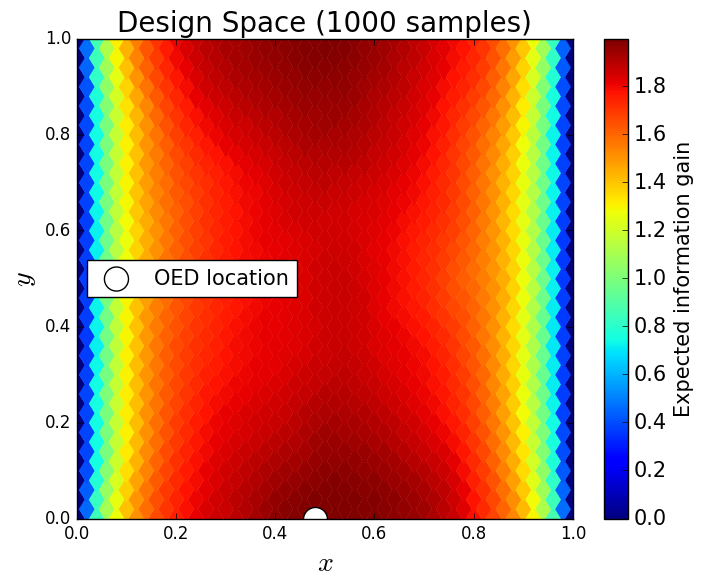}}
    	\scalebox{0.4}{\includegraphics{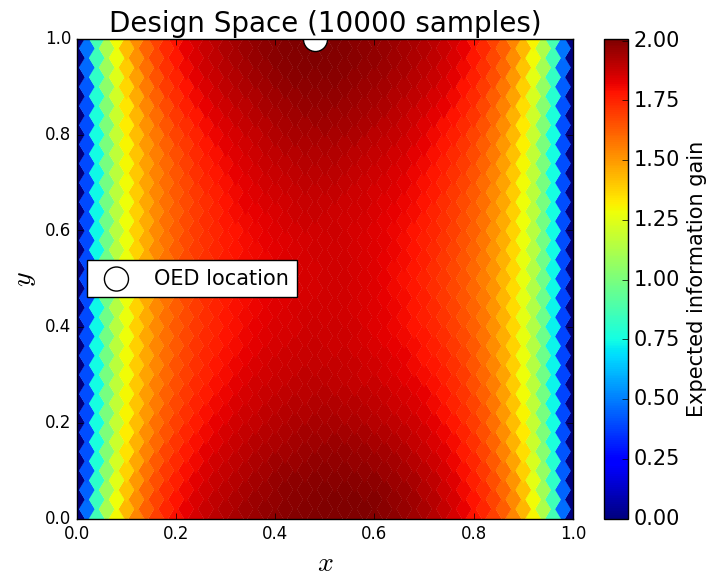}}
\caption{The expected information gain over the design space (which is $\Omega$ in this example) approximated using 50, 100, 1,000 and 10,000 samples from the prior.  Notice the small changes in the design space as we increase the number of samples from 1,000 to 10,000.  This suggests we compute accurate approximations to the design space using as few as 1,000 model evaluations.}
\label{fig:porousmedia}
\end{figure}

\begin{table}
\centering
\begin{tabular}[ht]{ccccc}
\hline
Design Location & 50 & 100 & 1,000 & 10,000 \\ \hline \hline
$(0.48, 1)$ & 1.966 & 2.001 & 1.992 & 2.008 \\
$(0.5, 0.98)$ & 1.952 & 1.963 & 1.993 & 2.007 \\
$(0.46, 0.98)$ & 1.930 & 1.985 & 1.986 & 2.006 \\
$(0.52, 0)$ & 1.777 & 1.915 & 1.999 & 2.006 \\
$(0.56, 0)$ & 1.751 & 1.863 & 1.996 & 2.006 \\ \hline
\end{tabular}
\caption{The top 5 experimental designs chosen using the full set of 10,000 samples.  For each of these designs, we compute $E(I_{Q^z})$ for 50, 100, 1,000 and 10,000 samples.  Notice the change in $E(I_{Q^z})$ for a given design decreases as we increase to 10,000 samples.}
\label{tab:porousmedia}
\end{table}

Next, we consider the greedy OED problem of placing 8 sensors within the physical domain.  We choose to use all of the available 10,000 samples to solve this problem. In Figure~\ref{fig:porousmedia_greedy}, we see the design space as a function of the previously determined locations of placed sensors.  We observe a strong symmetry to this problem, as is expected due to the symmetry of the physical process defined on this domain with the given boundary conditions.  In the bottom right of Figure~\ref{fig:porousmedia_greedy}, notice the very small range of the color bar indicating the possible values of the expected information gain.  This suggests, for this example, we expect there is a limit on the number of useful sensor locations for informing likely parameter values.

\begin{figure}
    \centering
    \scalebox{0.38}{\includegraphics{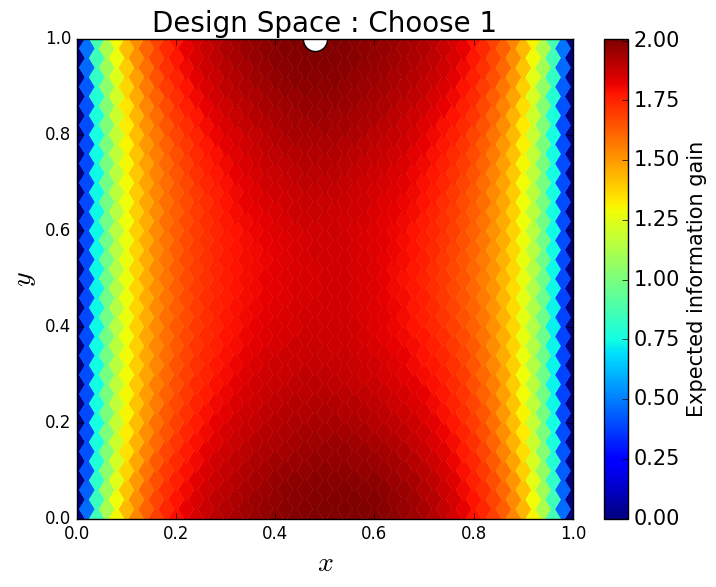}}
    	\scalebox{0.38}{\includegraphics{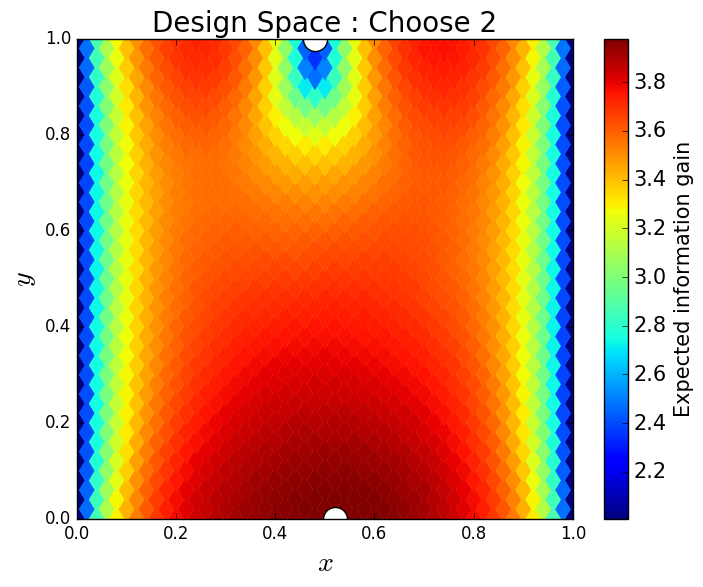}}\\
    \scalebox{0.38}{\includegraphics{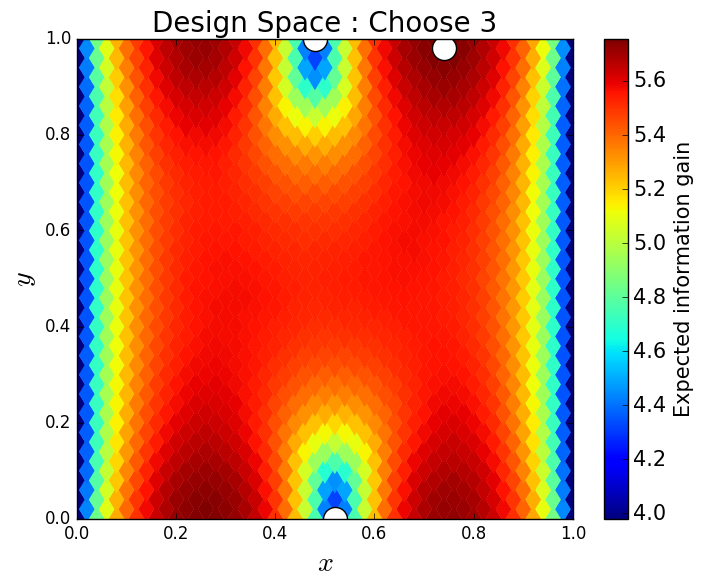}}
    	\scalebox{0.38}{\includegraphics{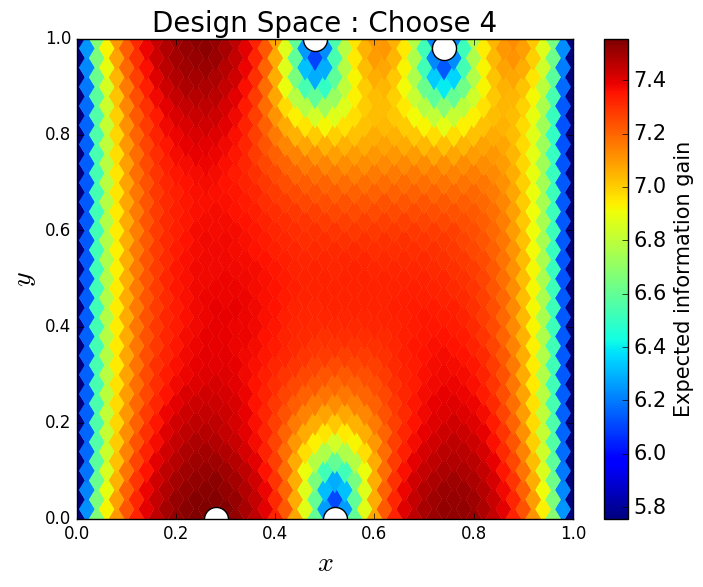}}
    \scalebox{0.38}{\includegraphics{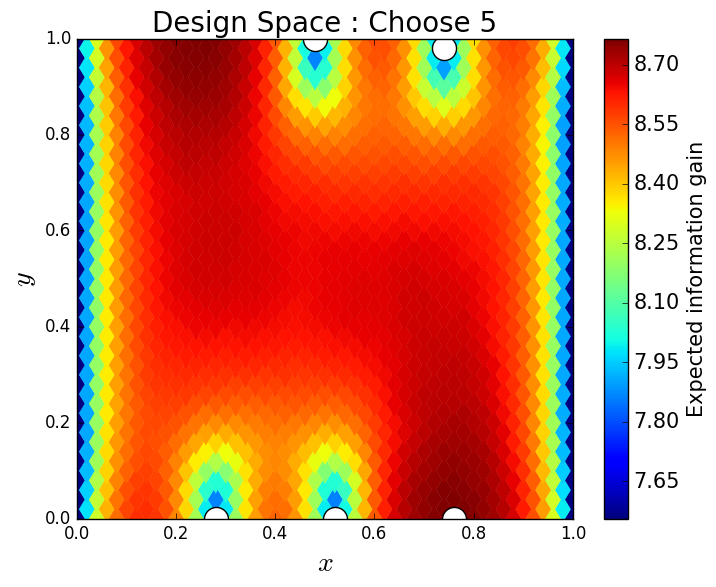}}
    	\scalebox{0.38}{\includegraphics{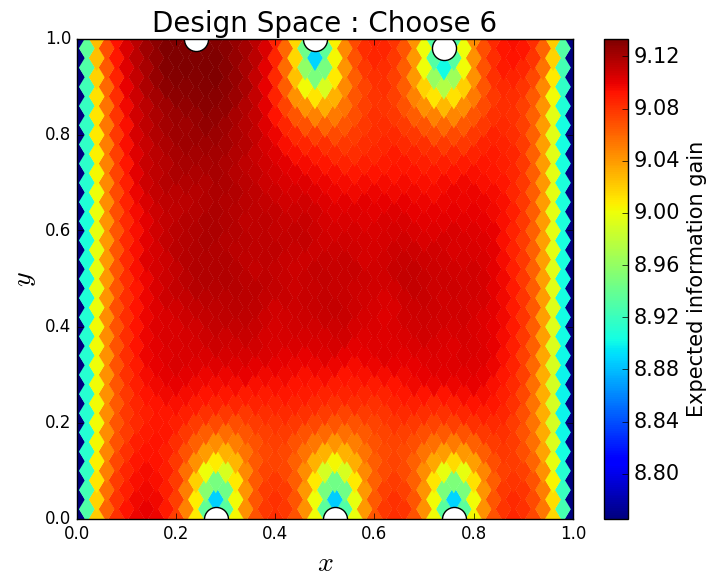}}\\
    \scalebox{0.38}{\includegraphics{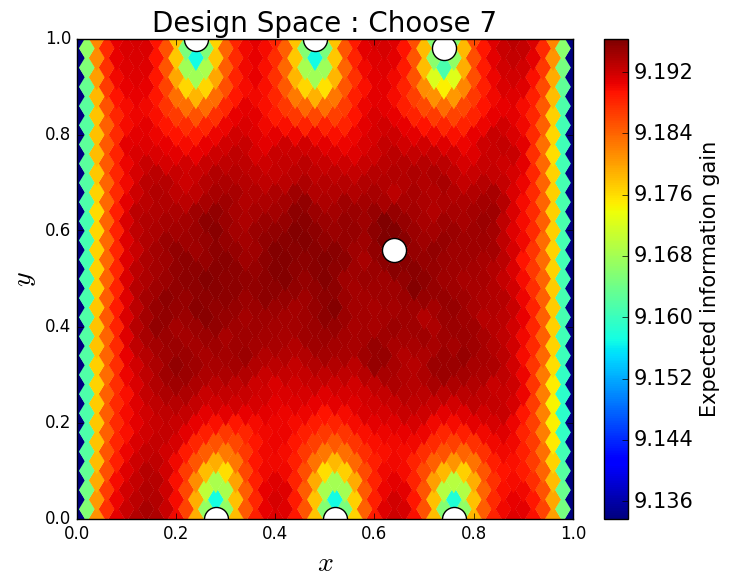}}
    	\scalebox{0.38}{\includegraphics{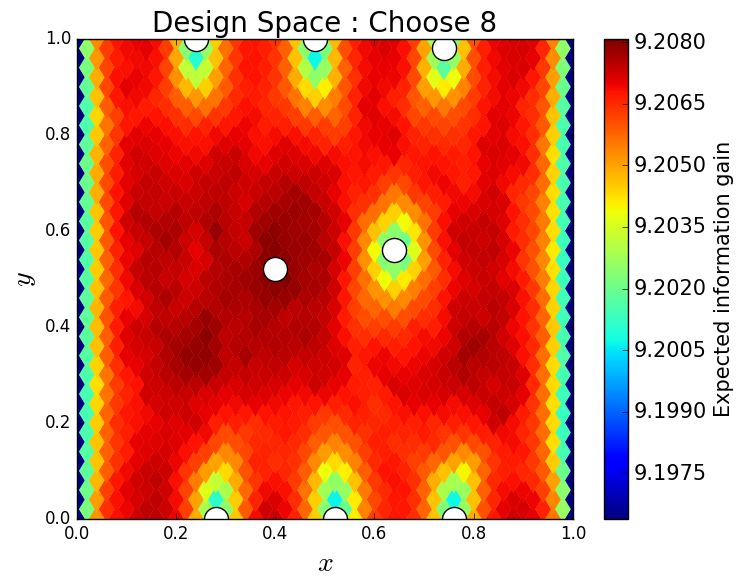}}
\caption{The expected information gain over the design space as a function of previously chosen sensor locations.  Note that the range of the color bar changes in the progression of the figures.  In the bottom right, we see the greedy optimal location of eight sensors within the physical domain.}
\label{fig:porousmedia_greedy}
\end{figure}

Lastly, we consider the exhaustive OED problem of placing 5 sensors within the physical domain and, for computational feasibility, restrict the possible locations of these 5 sensors to 25 points in the physical domain, see Figure~\ref{fig:porousmedia_exhaustive}.
We choose to use 1,000 samples to solve this problem. In Figure~\ref{fig:porousmedia_exhaustive}, we plot the design space for a single sensor location using these 1,000 samples and show the optimal location of 1, 2, 3, 4 and 5 sensors.
The results are quite similar to the greedy results previously described.

\begin{figure}
    \centering
    \scalebox{0.4}{\includegraphics{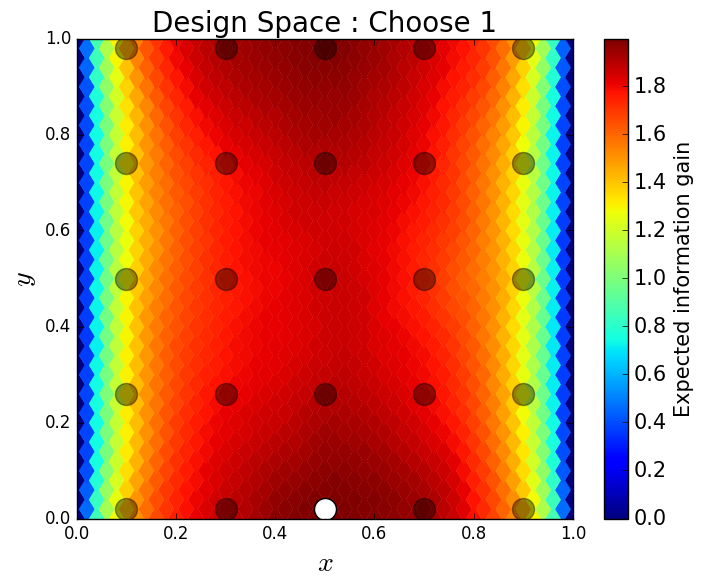}}
    	\scalebox{0.4}{\includegraphics{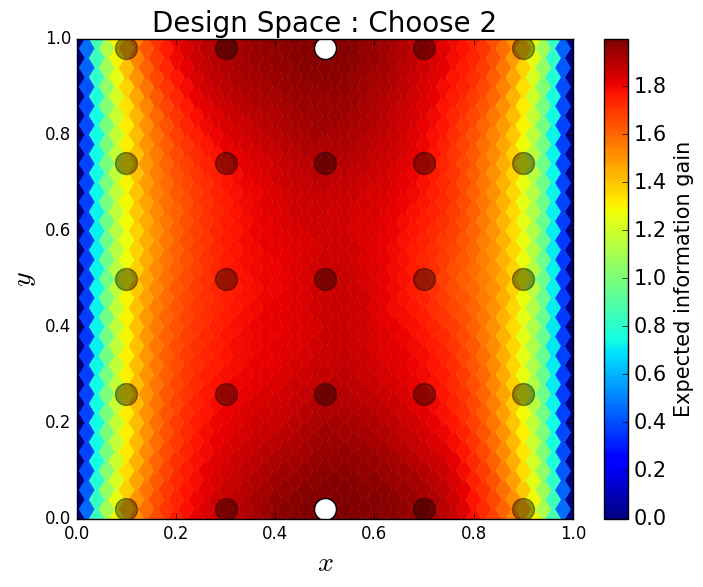}}\\
    \scalebox{0.4}{\includegraphics{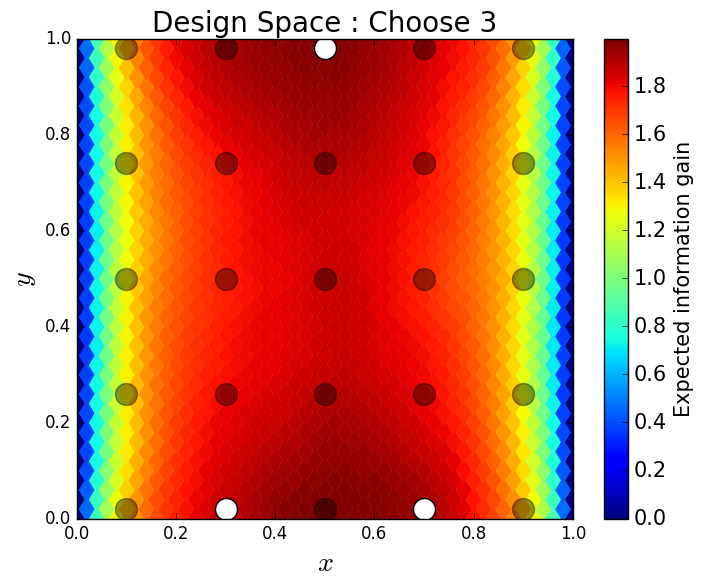}}
    	\scalebox{0.4}{\includegraphics{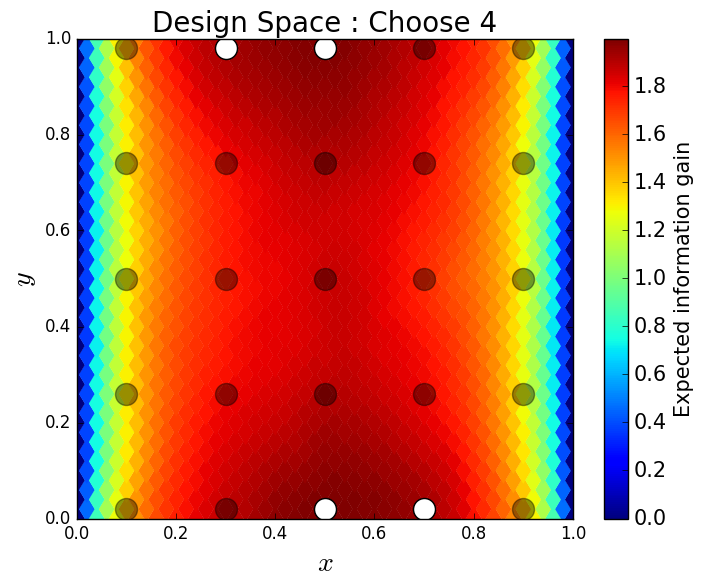}}
    \scalebox{0.4}{\includegraphics{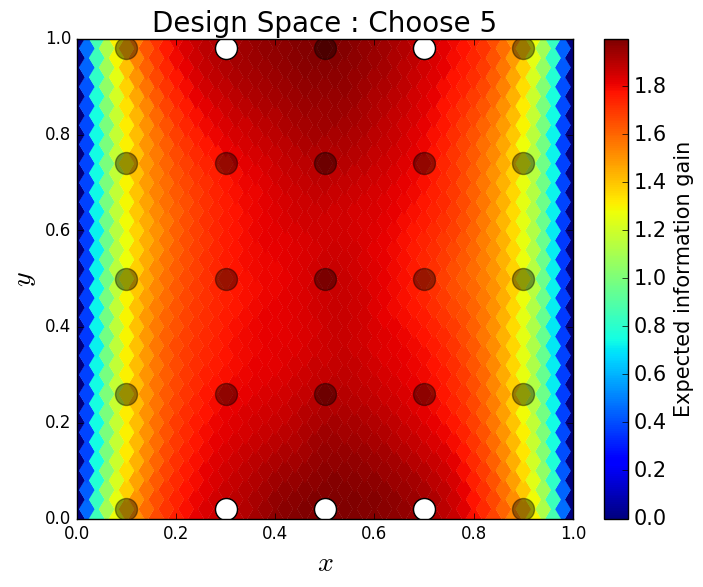}}
\caption{In black, we show the possible locations of the sensors and in white we show the optimal location(s) for 1, 2, 3, 4 and 5 sensors.  In the bottom, we see the optimal location of five sensors within the physical domain.  Note that each color bar is indicative of the expected information gain for the first sensor location, not for the expected information gain for multiple sensors.}
\label{fig:porousmedia_exhaustive}
\end{figure}

%%% Local Variables:
%%% mode: latex
%%% TeX-master: "../oed_cbayes"
%%% End:

%%%%%%%%%%%%%%%%%%%%%%%%%%%%%%%%%%%%%%%%%%%%%%%
%%%%%%%%%%%%%%%%%%%%%%%%%%%%%%%%%%%%%%%%%%%%%%%
\section{Conclusion} \label{sec:conclusion}
In this manuscript, we developed an OED formulation based on the recently developed
{\em consistent} Bayesian approach for solving stochastic inverse
problems. We used the Kullback-Leibler divergence and the
posterior obtained using consistent Bayesian inference to measure
the information gain of a design and present a discrete optimization
procedure for choosing the optimal experimental design that maximizes the
expected information gain.
The optimization procedure presented in this paper is limited in terms of the
number of observations we can consider, but was chosen to focus attention on the
definition and approximation of the expected information gained.
More efficient strategies, utilizing gradient-based methods on continuous design spaces,
will be pursued in a future work.
We discussed a characterization of the space of observed densities
needed to compute the expected information gain and
a computationally efficient approach for rescaling observed densities
to satisfy the requirements of the consistent Bayesian approach.
Numerical examples were given to highlight the properties and
utility of our approach.

\section{Acknowledgments}
J.D.~Jakeman's work was supported by DARPA EQUIPS. Sandia National
Laboratories is a multimission laboratory managed and operated by
National Technology and Engineering Solutions of Sandia, LLC., a
wholly owned subsidiary of Honeywell International, Inc., for the
U.S. Department of Energy's National Nuclear Security Administration
under contract DE-NA-0003525.
\bibliographystyle{plain}
\bibliography{oed_cbayes}

\end{document}